\documentclass[prd,reprint,10pt,twoside,tightenlines,nofootinbib,showpacs,notitlepage,superscriptaddress,aps]{revtex4-2}

\pdfoutput=1
\usepackage{dcolumn}
\usepackage{amsmath,amsfonts,amssymb,latexsym}
\usepackage[sort&compress]{natbib}
\usepackage{grffile,epstopdf}
\usepackage{graphicx}
\usepackage{hhline}
\usepackage{xcolor}
\usepackage{hhline}
\usepackage{physics}
\usepackage[T1]{fontenc}
\usepackage{selinput}
\usepackage{babel}

\makeindex

\begin{document}

\title{Transport properties and equation-of-state of hot and dense QGP matter near the critical end-point in the phenomenological dynamical quasiparticle model.}
		
\author{Olga Soloveva}
\email{soloveva@itp.uni-frankfurt.de} 
\address{Helmholtz Research Academy Hesse for FAIR (HFHF), GSI Helmholtz Center for Heavy Ion Physics, Campus Frankfurt, 60438 Frankfurt, Germany}
\address{Institut f\"ur Theoretische Physik, Johann Wolfgang Goethe-Universit\"at,
Max-von-Laue-Str.\ 1, D-60438 Frankfurt am Main, Germany}

\author{J\"org Aichelin} 
\address{SUBATECH,  University  of  Nantes,  IMT  Atlantique,IN2P3/CNRS  4  rue  Alfred  Kastler,  44307  Nantes  cedex  3,  France}
\affiliation{Frankfurt Institute for Advanced Studies, Ruth Moufang Str. 1, 60438 Frankfurt, Germany}

\author{Elena Bratkovskaya}
\address{GSI Helmholtzzentrum f\"ur Schwerionenforschung GmbH,
Planckstrasse 1, D-64291 Darmstadt, Germany}
\address{Institut f\"ur Theoretische Physik, Johann Wolfgang Goethe-Universit\"at,
Max-von-Laue-Str.\ 1, D-60438 Frankfurt am Main, Germany}
\address{Helmholtz Research Academy Hesse for FAIR (HFHF), GSI Helmholtz Center for Heavy Ion Physics, Campus Frankfurt, 60438 Frankfurt, Germany}

\date{\today }

\begin{abstract}
We extend the effective dynamical quasiparticle model (DQPM) - constructed for the description of non-perturbative QCD phenomena of the strongly interacting quark-gluon plasma (QGP) - to large baryon chemical potentials, $\mu_B$, including a critical end-point and a 1st order phase transition. 
The DQPM description of quarks and gluons is based on partonic propagators with complex selfenergies where the real part of the selfenergies is related to the quasiparticle mass and the imaginary part to a finite width of their spectral functions (i.e. the imaginary parts of the propagators).
In DQPM the determination of complex selfenergies for the partonic degrees of freedom at zero and finite $\mu_B$ has been performed by adjusting the entropy density to the lattice QCD (lQCD) data. The temperature-dependent effective coupling (squared) $g^2(T/T_c)$, as well as the effective masses and widths or the partons are based on this adjustment. 
The novel extended dynamical quasiparticle model, named "DQPM-CP", makes it possible to describe thermodynamical and transport properties of quarks and gluons in a wide range of temperature, $T$, and baryon chemical potential, $\mu_B$, and reproduces the equation-of-state (EoS) of lattice QCD calculations in the crossover region of finite $T, \mu_B$. 
We apply a scaling ansatz for the strong coupling constant near the critical endpoint CEP, located at ($T^{CEP}$,
$\mu^{CEP}_B) = (0.100, 0.960)$ GeV. We show the equation-of-state as well as the speed of sound for $T>T_c$ and for a wide range of $\mu_B$, which can be of interest for hydrodynamical simulations. 
Furthermore, we consider two settings for the strange quark chemical potentials (I) $\mu_q=\mu_u=\mu_s=\mu_B/3$ and (II) $\mu_s=0,\mu_u=\mu_d=\mu_B/3$. The isentropic trajectories of the QGP matter are compared for these two cases. 
The phase diagram of DQPM-CP is 
close to
PNJL calculations. The leading order pQCD transport coefficients of both approaches differ. This elucidates that the knowledge of the phase diagram alone is not sufficient to describe the dynamical evolution of strongly interacting matter.  
\end{abstract}
\maketitle

\section{Introduction}

The extension of the QCD phase diagram to a finite baryon chemical potential is a challenging task. It is believed that QCD matter undergoes a phase transition from the confined hadronic phase to the deconfined QGP phase if one increases the chemical potential at moderate temperatures and the transition line in ($T$, $\mu_B$) is expected to terminate at a critical end-point. This is the least explored area of the QCD phase diagram
but of particular interest for future experimental programs and theoretical studies (see recent review \cite{An:2021wof}). To realize
these studies in a viscous hydrodynamical model one has to know the equation-of-state, EoS, of strongly interacting matter but also the transport coefficients. The time evolution of the QGP medium, produced in heavy-ion collisions (HICs), can also be addressed in microscopic transport approaches, which provide the time evolution of the degrees of freedom of the system. They require in addition the knowledge of the microscopic properties of the partonic degrees of freedom, such as effective masses, widths and cross sections, which all may depend on $\mu_B$ and $T$. The large value of the running coupling requires 
nonperturbative methods as lattice QCD (lQCD) calculations, or effective models with a phenomenological input. Therefore, it is notoriously difficult to estimate thermodynamic properties of the deconfined QCD matter, especially in the vicinity of a phase transition.  

The lQCD calculations at vanishing baryon chemical potential are well established. However, due to the fermion sign problem, it is not at all easy to extend these calculations to a large 
baryon chemical potential. One possibility for exploring thermodynamic functions at $\mu_B > 0$ is to employ a Taylor expansion of the partition function in the vicinity of $\mu_B$ = 0. It shares with other approaches, which were designed for $\mu_B = 0$, that the uncertainty of the predictions increases with the increase of $\mu_B$.  

Here we present a new phenomenological model, the generalized quasiparticle model, DQPM-CP, for the  description of non-perturbative features of the (strongly interacting) QCD. It reproduces the lQCD EoS for $\mu_B = 0$ as well as the first coefficient of the Taylor expansion towards finite $\mu_B$ but can be extended to a wide range of $\mu_B$. For this we combine the findings of DQPM, with a new parametrisation of the coupling constant. One of the important features of the DQPM-CP is the appearance of 'critical' scaling in the vicinity of the CEP. The main goal of the DQPM-CP is to provide the microscopic and macroscopic properties of the partonic degrees of freedom  for the region of the phase diagram which is characterized by moderate $T$ and moderate or high $\mu_B$. Their knowledge allows subsequently to calculate the transport coefficients as well as the EoS, the ingredients of viscous hydrodynamic calculations.

 In the present study we employ results from different methods such as lQCD calculations and results from the $N_f=3$ PNJL model extended beyond mean field, because more rigorous approaches, such as Dyson-Schwinger equations (DSE) \cite{Fischer:2018sdj}, functional renormalization group FRG \cite{Braun:2020ada} or pQCD/HTL calculations \cite{Ghiglieri:2020dpq} can presently not cover the interesting observables in the full ($T$, $\mu_B$)- plane. There are no fully consistent calculations within a single approach, which includes the QGP thermodynamic observables and simultaneously the transport coefficients for the region of moderate and high $\mu_B$ or $\mu_q$. The presented results are model-dependent, however, qualitatively in agreement with the results from various effective models such as PNJL, NJL, LSM, and non-conformal holographic models. Only at moderate baryon density we can compare to the more rigorous methods such as lQCD, FRG, DSE.
Nevertheless, the demand for an EoS and transport coefficients at moderate and high baryon chemical potentials, is high due to the ongoing investigation of heavy-ion collisions (HIC) by  transport or/and hydrodynamic models \cite{An:2021wof}.  These investigations aim at the exploration of observables, which may carry information about this region of the phase diagrams. Therefore the presented results , even if they allow only for qualitative predictions, can be useful for  transport studies, which have multiple issues to solve in the region of moderate baryon density while awaiting results from more rigorous approaches. 

The main advantage of the use of quasiparticle models is the simple implementation in the transport framework for the evolution of the QGP matter. The DQPM has been implemented in the PHSD transport approach  \cite{Cassing:2009vt,Moreau:2019vhw}, whereas the QPM \cite{Plumari:2011mk} is implemented in the Catania transport approach \cite{Scardina:2017ipo}. Recently also results from approximate models of QCD, like that from NJL and PNJL models, have been implemented in the AMPT model \cite{Sun:2020uoj} via scalar and vector potentials. The goal of the presented model is to interpolate the EoS and partonic properties such as effective masses, scalar potential and cross-sections between the region of high $T$ and $\mu_B=0$ to the region of moderate $T$ and high $\mu_B$. In particular, we are currently working on the implementation of the DQPM-CP in the PHSD transport approach \cite{Cassing:2009vt,Moreau:2019vhw}.

The degrees of freedom of the DQPM are strongly interacting dynamical quasiparticles -- quarks and gluons -- with a broad spectral function, whose 'thermal' masses and widths increase with growing temperature. The knowledge of the $T$ and $\mu_B$ dependence of the mass of our degrees of freedom allows for the calculation of transport coefficients in lowest order in pQCD. They can be compared with the transport coefficients,  calculated recently in the PNJL approach, which has a very similar phase diagram, but other degrees of freedom (interacting massless quarks and no gluons). The comparison of the transport coefficients shows that they depend indeed on the properties of the degrees of freedom and may be rather different in two theories with almost the same phase diagram.  \\
The paper is organized as follows: In Sec. \ref{sec2} we give a brief review of the basic ingredients of the dynamical quasiparticle model and its extension to the finite $\mu_B$ region. In Sec. \ref{sec3} we discuss the thermodynamic observables for two setups of quark chemical potential: (I) $\mu_q=\mu_u=\mu_s=\mu_B/3$ and (II) $\mu_s=0,\mu_u=\mu_d=\mu_B/3$. Furthermore, we study second order derivatives of the partition function, such as speed of sound and specific heat, and isentropic trajectories of the QGP matter. 
Further in Sec. \ref{sec4} we present transport coefficients of the DQPM-CP such as the specific shear viscosity and ratio of electric $\sigma_{QQ}/T$, baryon  $\sigma_{BB}/T$ and strange $\sigma_{SS}/T$ conductivities to temperature based on the relaxation time approximation of the Boltzmann equation. In addition, we show the ratio of dimensionless transport coefficients for the full range of chemical potentials. We finalize our study with conclusions in Sec. \ref{sec5}.

\section{\label{sec2}Basic properties of the quasiparticle model }

\subsection{Main ingredients of the off-shell quasiparticle models}
\begin{figure}[!ht]
 \centering
\begin{minipage}[h]{1\linewidth}
\center{\includegraphics[width=0.96\linewidth]{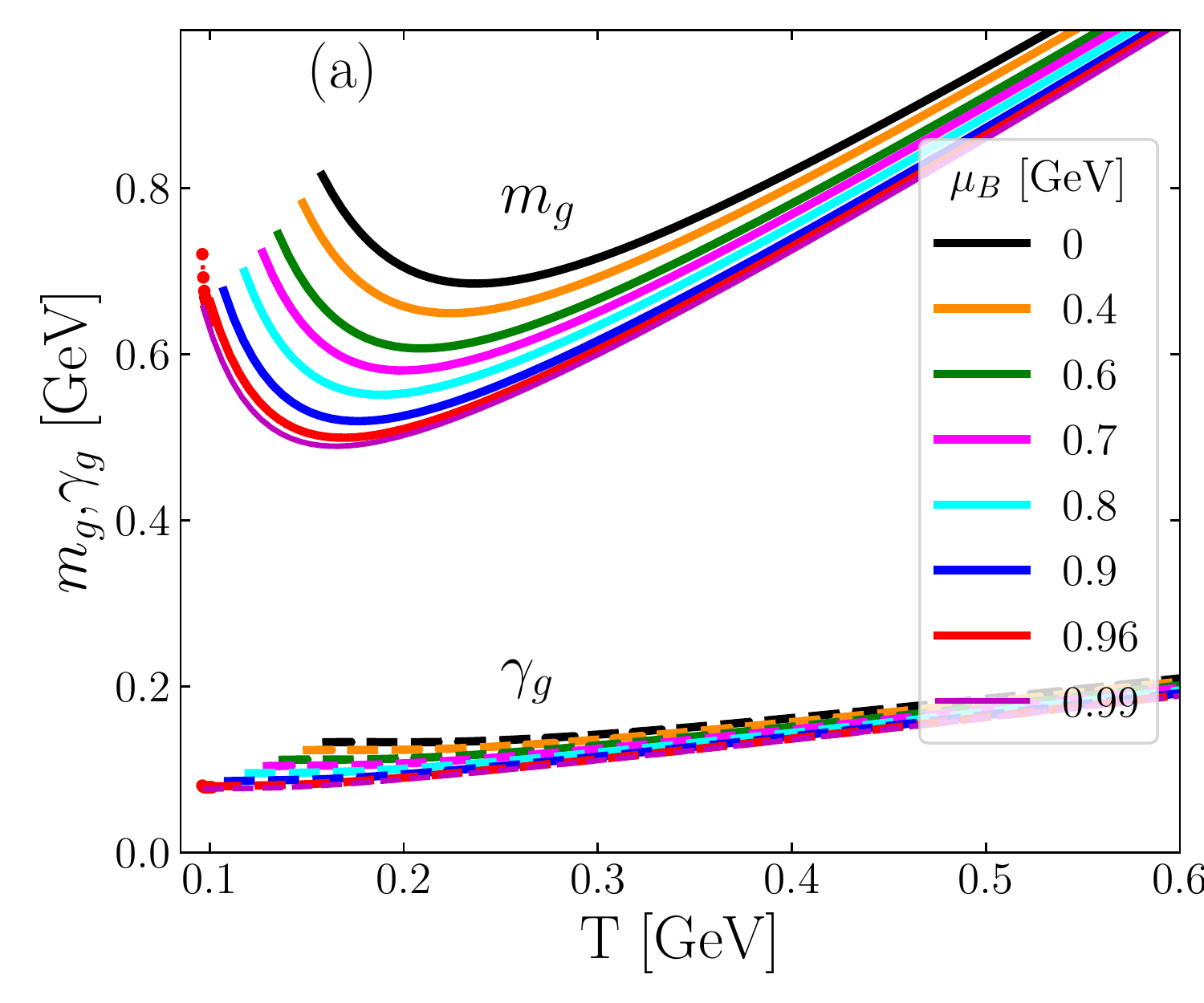}} 
\end{minipage}
\begin{minipage}[h]{1\linewidth}
\center{\includegraphics[width=0.96\linewidth]{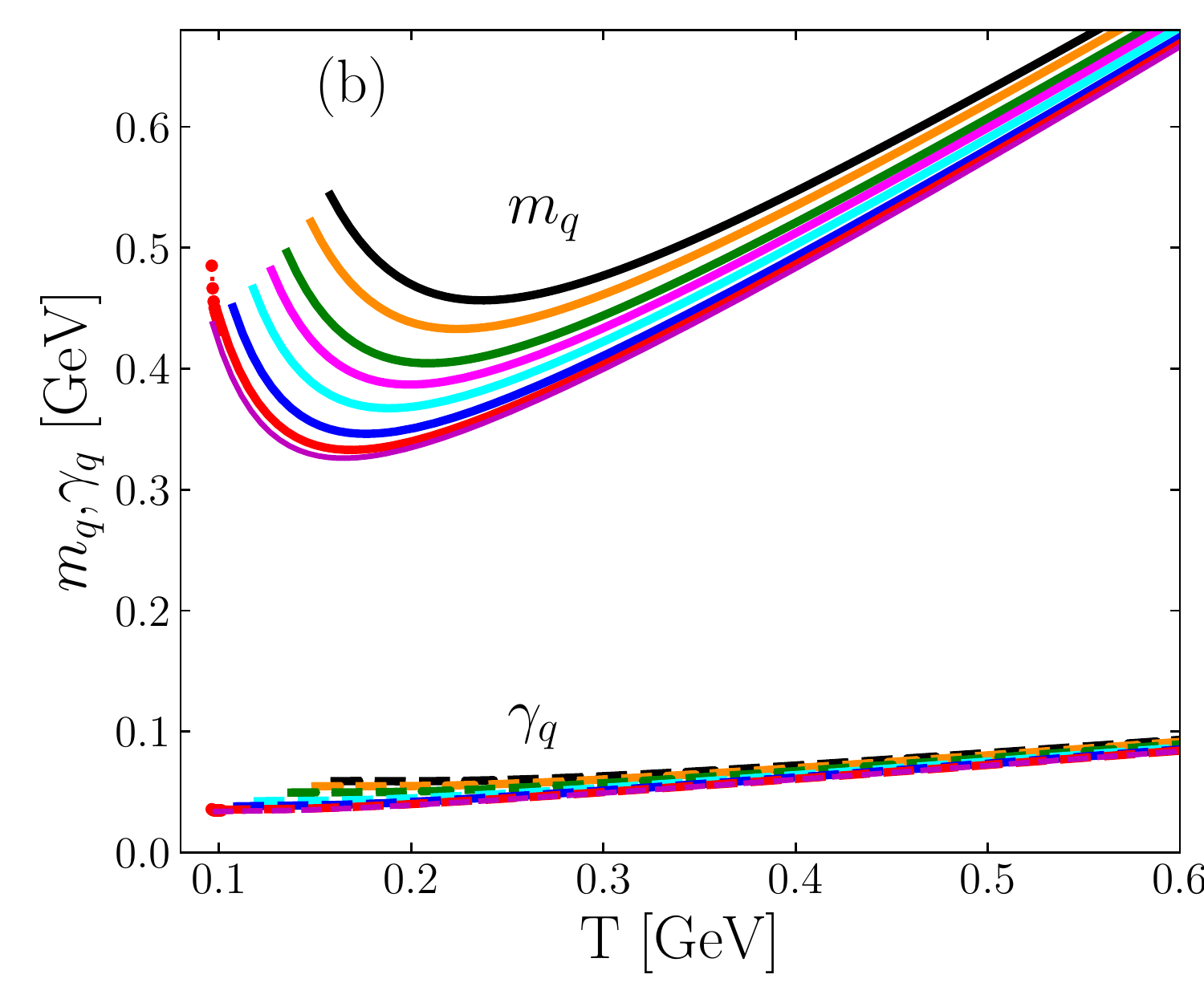}} 
\end{minipage}
  \caption{The DQPM-CP pole masses and widths for the gluons (a) and light quarks (b) given by Eqs. (\ref{polemass_g_dqpm}-\ref{eq:widths}) as a function of temperature in region $T>T_c(\mu_B)$ for fixed baryon potentials $\mu_B\geq 0$. }
  \label{fig:masses_mu}
    \end{figure}

In the dynamical quasiparticle model, DQPM
\cite{Peshier:2005pp,Cassing:2007nb,Cassing:2007yg,Moreau:2019vhw}, 
the QGP medium is described in terms of strongly interacting quasiparticles, quarks and gluons. 
The quasiparticles are massive and can be characterized by broad spectral functions  $\rho_i$ ($i=q, {\bar q}, g$), which are no
longer $\delta$-functions in the invariant mass squared but given by
	\begin{align}
		\rho_i(\omega,{\bf p}) = &
		\frac{\gamma_i}{\tilde{E}_{i,\mathbf{p}}} \left(
		\frac{1}{(\omega-\tilde{E}_{i,\mathbf{p}})^2+\gamma_i^2} - \frac{1}{(\omega+\tilde{E}_{i,\mathbf{p}})^2+\gamma_i^2}
		\right)\  \nonumber \\
		& = \frac{4\omega\gamma_i}{\left( \omega^2 - \mathbf{p}^2 - m^2_i \right)^2 + 4\gamma^2_i \omega^2}
		\label{eq:rho}.
	\end{align}
	Here, we introduced the off-shell energy $\tilde{E}_{i,\mathbf{p}} = \sqrt{ {\bf p}^2+m_i^2-\gamma_i^2 }$, with $m_i$,$\gamma_i$ are being the pole mass and width,
which differ for quarks, antiquarks and gluons. Then the quasiparticle (retarded) propagators can be expressed in the Lehmann representation via the spectral function:
\begin{equation}
\Delta_i(\omega,{\bf p})=  \int_{-\infty}^{\infty} \frac{d\omega'}{2\pi}\frac{\rho_i(\omega',{\bf p})}{\omega-\omega'}=\frac{1}{\omega^2 -{\bf p}^2 -m_i^2 +2 i\gamma_i\omega }
\end{equation}

The quasiparticle pole masses for gluons and quarks are
defined, inspired  by the asymptotic HTL masses
\cite{Bellac:2011kqa,Linnyk:2015rco}, by
	\begin{equation}
	 m^2_{g}(T,\mu_\mathrm{B})=C_g \frac{g^2(T,\mu_\mathrm{B})}{6}T^2\left(1+\frac{N_{f}}{2N_{c}}
	+\frac{1}{2}\frac{\sum_{q}\mu^{2}_{q}}{T^2\pi^2}\right)
		  \label{polemass_g_dqpm}
	\end{equation}
	
	\begin{equation}
	  m^2_{q(\bar q)}(T,\mu_\mathrm{B})=C_q \frac{g^2(T,\mu_\mathrm{B})}{4}T^2\left(1+ \frac{\mu^{2}_{q}}{T^2\pi^2}\right)
	  \label{polemass_q_dqpm},
	\end{equation}
	where $N_{c}=3$ and $N_{f}=3$ denote the number of colors and the number of flavors, respectively. $C_q = \frac{N_c^2 - 1}{2 N_c} = 4/3$ and $C_g = N_c = 3$ are the QCD color factors for quarks and for gluons, respectively. The strange quark has a larger bare mass which needs to be considered in its dynamical quasi particle pole mass. We fix $m_s(T,\mu_\mathrm{B})= m_{u}(T,\mu_\mathrm{B})+ \Delta m$ and $\Delta m \approx$ 30 MeV \cite{Moreau:2019vhw}.	

	Furthermore, the quasiparticles in DQPM have thermal widths, which are adopted in the form \cite{Berrehrah:2016vzw,Linnyk:2015rco}
	\begin{equation}
		\gamma_{j}(T,\mu_\mathrm{B}) = \frac{1}{3} C_j \frac{g^2(T,\mu_\mathrm{B})T}{8\pi}\ln\left(\frac{2c_m}{g^2(T,\mu_\mathrm{B})}+1\right).
	\label{eq:widths}
	\end{equation}
The parameter $c_m$, which is related to a magnetic cut-off, is fixed to $c_m = 14.4$.\\
In the DQPM the coupling constant at $\mu_B =0$ is parameterized employing the entropy density $s(T,\mu_\mathrm{B} = 0)$ from lattice QCD calculations of Refs. \cite{Borsanyi:2012cr,Borsanyi:2013bia} in the following way:
	\begin{equation}
	g^2(T,\mu_\mathrm{B} = 0) = d \Big( \left(s(T,0)/s^\mathrm{QCD}_{\mathrm{SB}} \right)^e -1 \Big)^f,
	\label{coupling_DQPM}
	\end{equation}
with the Stefan-Boltzmann entropy density $s_{\mathrm{SB}}^{\mathrm{QCD}}/T^3 = 19/9\ \pi^2 $ and the dimensionless parameters $d = 169.934$, $e = -0.178434$ and $f = 1.14631$.

We note that the DQPM has been used to explore the crossover region in the phase diagram by introducing an effective coupling constant which depends on the baryon chemical potential. In this region of a moderate baryon chemical potential the basic thermodynamic observables, computed in lQCD, show a smooth $\mu_B$ dependence. Therefore we expect a similar behaviour for the effective coupling. 

The effective coupling at finite baryon chemical potential $\mu_\mathrm{B}$ is obtained by applying the 'scaling hypothesis' introduced in \cite{Cassing:2007nb}. It assumes that $g^2$ is a function of the ratio of the effective temperature
	 \begin{equation}
	 T^* = \sqrt{T^2+\mu^2_q/\pi^2}
	 \label{eq:tstar}
	 \end{equation}
 (where the quark chemical potential is defined as $\mu_q=\mu_u=\mu_s=\mu_B/3$ ) and the $\mu_\mathrm{B}$-dependent critical temperature $T_c(\mu_\mathrm{B})$ as \cite{Berrehrah:2016vzw}:
	 	\begin{equation}
	    T_c(\mu_B) = T_c(0) (1-\alpha \mu_B^2)^{1/2},
	    \label{eq:dqpmTc}
	\end{equation}
where $T_c(0)$ is the critical temperature at vanishing chemical potential ($\approx 0.158$ GeV) and $\alpha = 0.974$ GeV$^{-2}$.
Thus, the DQPM effective coupling $\alpha_S^{DQPM}(T,\mu_\mathrm{B})$ reads

\begin{align}
\alpha_S^{DQPM}(T,\mu_\mathrm{B}) \equiv \begin{cases}
& \mu_\mathrm{B} = 0: \ g^2(T,\mu_\mathrm{B} = 0)/(4\pi)  \\
& \mu_\mathrm{B} > 0: \ g^2(T_{scale}(T,\mu_\mathrm{B}))/(4\pi),\\
& \rm{with} \ \ T_{scale} = \frac{T^*}{T_c(\mu_\mathrm{B})/T_c(0)} . 
	\end{cases}
\end{align}\label{eq:as_DQPM}

Having fixed the quasiparticle properties (or propagators) as described above, one can 
evaluate the basic thermodynamic observables: the entropy density $s(T,\mu_B)$, the pressure $P(T,\mu_B)$ and energy
density $\epsilon(T,\mu_B)$ in a straight forward manner by starting with
the quasiparticle entropy density and number density. 	The entropy density and the quark number density follow from the same thermodynamic potential $\Omega[\Delta,S_q]$ \cite{Vanderheyden:1998ph,Blaizot:2000fc}, which is expressed as a functional of the full quasiparticle propagators for gluons and quarks($\Delta,S_q$) in a symmetry-conserving ('$\Phi$-derivable') two-loop approximation:
\begin{equation}
s^{dqp} =
\label{sdqp}
\end{equation}
\begin{align*}
  & -d_g \int \frac{d\omega}{2 \pi} \frac{d^3p}{(2 \pi)^3} \frac{\partial f_g}{\partial T} \left( \Im(\ln \Delta^{-1})- \Im \Pi \Re \Delta \right) \\
  & -d_q \int \frac{d\omega}{2 \pi} \frac{d^3p}{(2 \pi)^3} \frac{\partial f_q(\omega-\mu_q)}{\partial T} \left( \Im(\ln S_q^{-1})- \Im \Sigma_q \Re S_q \right) \\
  & -d_{\bar{q}} \int \frac{d\omega}{2 \pi} \frac{d^3p}{(2 \pi)^3} \frac{\partial f_{\bar q}(\omega+\mu_q)}{\partial T} \left( \Im(\ln S_{\bar{q}}^{-1})- \Im \Sigma_{\bar{q}} \Re S_{\bar{q}} \right)
\end{align*}

\begin{equation}
 n^{dqp} =
 \label{nbdqp}
\end{equation}
\begin{align}
  & -d_q \int \frac{d\omega}{2 \pi} \frac{d^3p}{(2 \pi)^3} \frac{\partial f_q(\omega-\mu_q)}{\partial \mu_q} \left( \Im(\ln S_q^{-1})- \Im \Sigma_q \Re S_q \right) \nonumber \\
  & -d_{\bar{q}} \int \frac{d\omega}{2 \pi} \frac{d^3p}{(2 \pi)^3} \frac{\partial f_{\bar q}(\omega+\mu_q)}{\partial \mu_q} \left( \Im(\ln S_{\bar{q}}^{-1})- \Im \Sigma_{\bar{q}} \Re S_{\bar{q}} \right) \nonumber
\end{align}
~\\
where $f_g(\omega)$ and
$f_q(\omega-\mu_q)$ denote the
Bose-Einstein and Fermi-Dirac distribution functions (see Eq. (\ref{eq:Equilibrium})), respectively, while $\Delta
=(p^2-\Pi)^{-1}$, $S_q = (p^2-\Sigma_q)^{-1}$ and $S_{\bar q} =
(p^2-\Sigma_{\bar q})^{-1}$ stand for the full (scalar)
quasiparticle propagator of gluons $g$, quarks $q$ and antiquarks
${\bar q}$.  In Eq. (\ref{sdqp})-(\ref{nbdqp}) we consider for simplicity scalar (retarded) quasiparticle
self-energies $\Pi=\Sigma =  \Sigma_q
\approx \Sigma_{\bar q}$, which are expressed via dynamical masses and widths as $\Pi = m_i^2 -2i\gamma_i \omega_i$, where, for the off-shell case, $\omega_i$ is an independent variable. Furthermore, the number of transverse gluonic degrees of freedom is $d_g=2 \times (N_c^2-1)$
while for the fermion degrees of freedom we use $d_q= 2 \times N_c$ and $d_{\bar{q}}= 2 \times N_c$. \\

\subsection{Extension of quasiparticle DQPM-CP effective coupling constant for the inclusion of the CEP}
\begin{figure}
\centering
\includegraphics[width=0.5\textwidth]{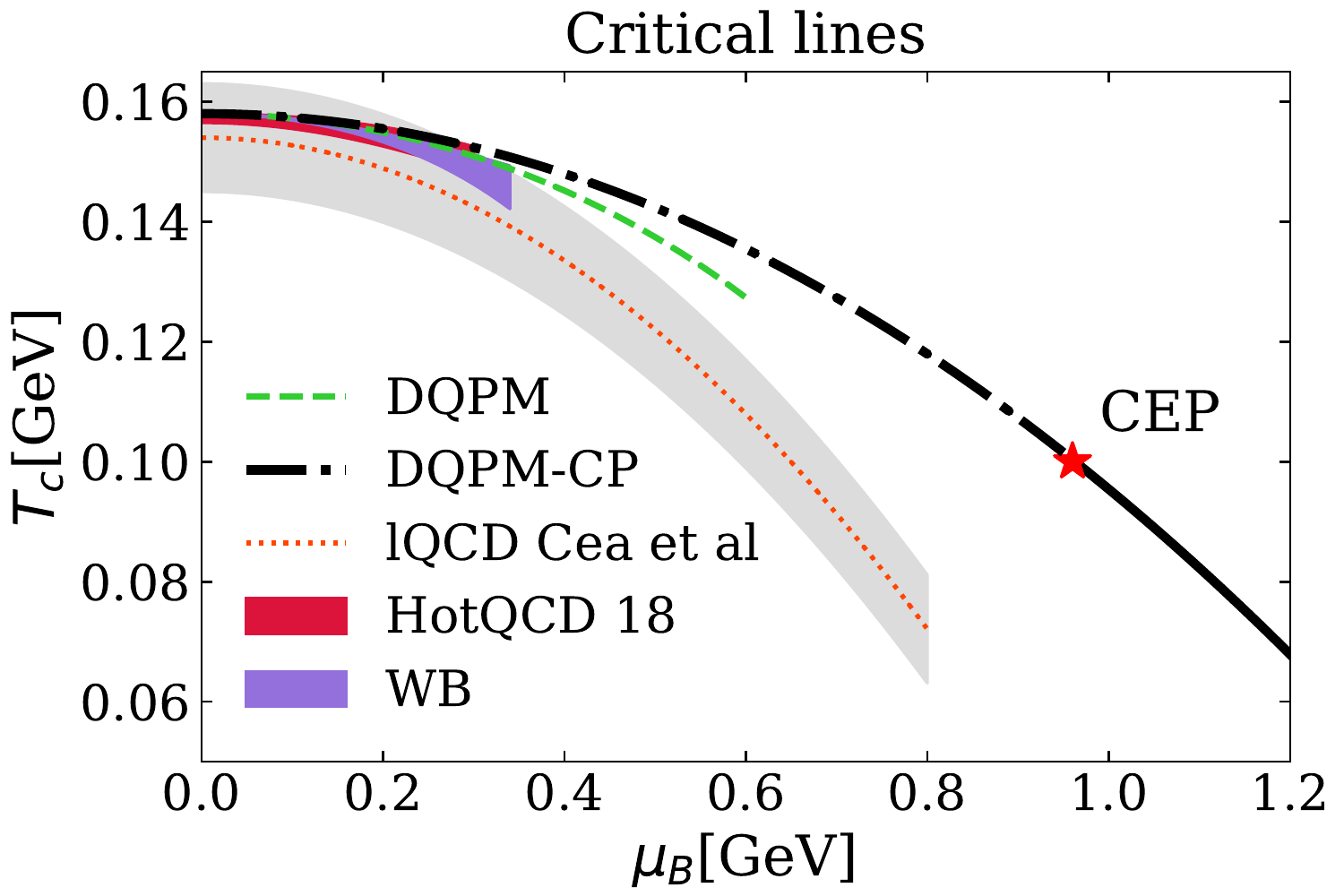}
\caption{\label{fig:Tc} Critical line of the DQPM-CP (black lines) and of the DQPM (green dashed line) in the $(T,\mu_B)$-plane of the QCD phase diagram. 
DQPM-CP ($\mu_q=\mu_s=\mu_B/3$): The finite temperature crossover (black dash-dotted line) at small chemical potential switches to a first-order transition (black solid line) at the hypothetical CEP (red star), which is located at $(0.10, 0.96)$ GeV.
DQPM ($\mu_q=\mu_s=\mu_B/3$): The finite temperature crossover (green dashed line).
Lines with colored areas represent lQCD estimates of $T_c(\mu_B)$ for QCD with $N_f= 2 + 1$: 
($\mu_q=\mu_s=\mu_B/3$) grey area - from Ref. \cite{Cea:2015cya}, ($\mu_q=\mu_s=\mu_B/3$) red area - from Ref. \cite{Borsanyi:2020fev}, ($\langle n_S \rangle=0$) violet area - from Ref. \cite{Bazavov:2018mes}. }
\end{figure}
Now we proceed with the extension of the DQPM to the region of large $\mu_B$
where a possible critical end-point is located. 
In order to extent the quasiparticle model to the large $\mu_B$ region 
and to describe the critical behavior near the CEP we depart from 
the 'scaling hypothesis', used for the moderate baryon chemical potentials 
in crossover region, and introduce
a simple parametrization of the coupling constant as a function of the scaled temperature and the baryon chemical potential.
To simplify an extension of the effective coupling for finite $\mu_B$ we parametrize $\alpha_s^{DQPM-CP}$ as a function of a dimensionless scaled temperature $x_T=T/T_c$. We determine first the parameters at vanishing quark or baryon chemical potential by fitting extracted values of $g^2_{DQPM}(T,\mu_B=0)$ from Eq. (\ref{coupling_DQPM}) as a function $f(T/T_c,\mu_B=0)$ ($T_c=0.158$) with help of the nonlinear least-squares (NLLS) Marquardt-Levenberg algorithm.
Later we use critical line values of $T_c$ for each value of baryon chemical potential. The critical line of the present model, which is an input parameter for our calculations, reads:
\begin{equation}
    T_c(\mu_B)=T_c(0)[1-\kappa_{PNJL}(\mu_B/T_c(0))^2],
    \label{eq:Tc_line}
\end{equation}
where $T_c(0)=0.158$ GeV is fixed in accordance with the results from lQCD \cite{Borsanyi:2020fev,Bazavov:2018mes}, while $\kappa_{PNJL}=0.00989$ corresponds to the estimates from the PNJL model \cite{Fuseau:2019zld}.

Figure \ref{fig:Tc} shows a comparison of the critical lines of the DQPM (green dashed line), of the DQPM-CP and the predictions from the lQCD calculations. The DQPM-CP phase boundary, given by Eq. (\ref{eq:Tc_line}), is shown as a black dashed-dotted line in the crossover region, means for moderate baryon chemical potentials. The critical endpoint in the presented model is located at ($T^{CEP}$,
$\mu^{CEP}_B) = (0.100, 0.960)$ GeV. 

The exact location of the CEP is an open question and there are many predictions from various methods (for a compilation of theoretical predictions for ($T^{CEP}$,
$\mu^{CEP}_B)$ we refer the reader to Fig.~6 in Ref.~\cite{Stephanov:2004wx}, and to Fig.~19 in Ref.~\cite{Guenther:2022wcr}). Current state-of-the-art lQCD results disfavor a critical point for $\mu_B/T\leq 3$ \cite{ Borsanyi:2012cr, HotQCD:2014kol,Bazavov:2018mes, Borsanyi:2021sxv}. Furthermore, it has been found that the temperature of the hypothetical chiral critical end point should not exceed the critical temperature of the chiral phase transition (for $m_u=m_d=0$) $T^{0}_c = 132^{+3}_{-6}$ MeV \cite{Karsch:2019mbv, HotQCD:2019xnw}. Recently, on approximate position of the chiral CEP (for vanishing external magnetic field) of $\mu^{CEP}_B = 0.800(0.140)$ GeV has been conjectured by the lQCD simulations of finite density QGP under external magnetic fields \cite{Braguta:2019yci}. It has been shown that the critical line of the PNJL model generally depends on the parameters of the model \cite{Costa:2009ae}. Many (P)NJL model predictions of the location of the CEP lie at high $\mu_B \sim 0. 8-1$ GeV ~\cite{Stephanov:2004wx}, for instance for the $N_f=2$ NJL model in Ref. \cite{Sasaki:2008um} ($T^{CEP}, \mu^{CEP}_B) = (0.081, 0.987)$ GeV ($\mu^{CEP}_B=3\mu^{CEP}_q$), while for the $N_f=3$ PNJL model considered in Ref. \cite{Motta:2020cbr} ($T^{CEP} ,\mu^{CEP}_B) = (0.121, 0.875)$ GeV.

We base this study on the predictions from the extended beyond mean field $N_f=3$ PNJL model  \cite{Fuseau:2019zld}. However, in order to fit lattice results at moderate $\mu_B \leq 0.6$ GeV we use the value of the pseudo-critical temperature at $\mu_B =0 $ from lQCD estimates. First, the value of the baryon chemical potential of the hypothetical CEP $\mu^{CEP}_B$ is chosen in accord with predictions from the PNJL model, then the temperature of the CEP follows from the chosen critical line $T_c(\mu_B)$. The chosen location of the hypothetical CEP is within the allowed range of the lQCD estimates. 
The first-order phase transition is shown as a solid black line. The DQPM phase boundary for moderate baryon chemical potentials, $\mu_B\leq 0.6$ GeV, given by Eq. (\ref{eq:dqpmTc}), is shown as a dashed green line. The colored areas illustrate the predictions from the lQCD calculations for QCD with $N_f= 2 + 1$: 
grey area - from Ref.  \cite{Cea:2015cya}, red area - from Ref. \cite{Borsanyi:2020fev}, violet area - from Ref. \cite{Bazavov:2018mes}.

To interpolate the EoS and microscopic properties of quarks and gluons between the region of the vanishing baryon chemical potential and the asymptotic behavior in the region of high baryon density $\mu_B \gg T$ ($T>T_c(\mu_B)$) we employ a simple ansatz for the $\mu_B$-dependence, which reflects the decrease of the effective coupling with $\mu_B$.
We assume that the coupling constant does not depend explicitly on the temperature but only on the scaled temperature ($x_T=T/T_c(\mu_B)$) ($T_c$ varies with $\mu_B$ according to the chosen critical line) for all $\mu_B\ge0$.
Furthermore, we introduce an additional factor $\sigma(\mu_B)(=1$ at $\mu_B=0$) to take into account the decrease of the coupling constant with $\mu_B$ at moderate $x_T$. At high $T$ ($T/T_c(\mu_B)\equiv x_T \gg 1$) the effective coupling constant is not affected by the baryon chemical potential. 
Therefore, the DQPM-CP coupling constant can be parametrized as a function of a scaled temperature and $\mu_B$:
\begin{figure}[h!]
\centering
\includegraphics[width=0.4\textwidth]{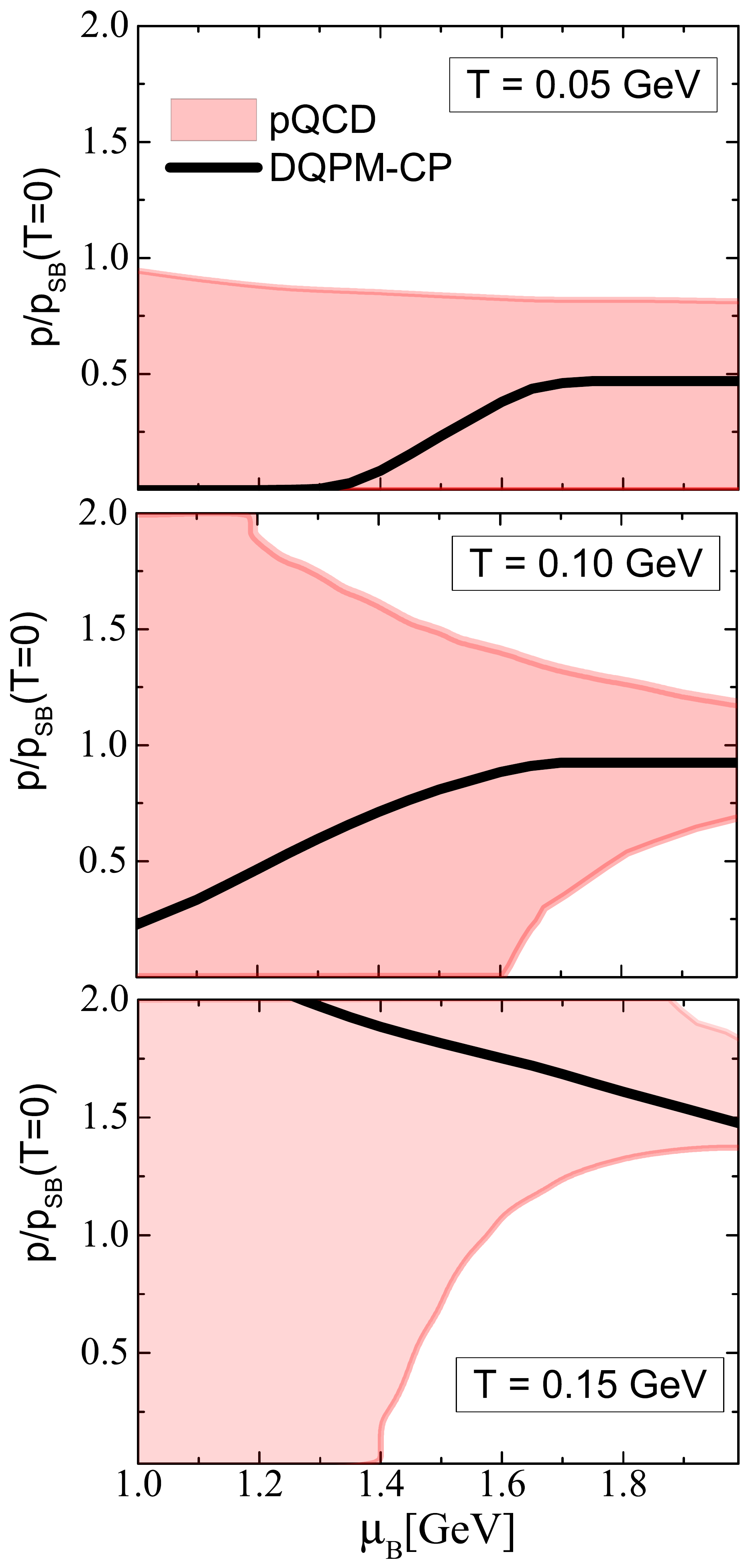}
\caption{\label{fig:p_pqcd}The scaled pressure $p/p_{SB}$ (black line) from the DQPM-CP in comparison to the pQCD results from Ref. \cite{Kurkela:2016was} (red area) as a function of baryon chemical potential for fixed temperatures $T = 0.05, 0.10, 0.150$ GeV. Scenario: $\mu_q=\mu_u=\mu_s=\mu_B/3$.}
\end{figure}
\begin{align}
  \alpha_{S}^{cross}= a_0 + \frac{a_2}{x_T^2} - \frac{a_3}{x_T^3} +  \frac{a_4}{x_T^4} + \frac{a_6 \cdot \sigma(\mu_B)}{x_T^6}.
  \label{eq:a_cross}
\end{align}
Here the coefficients $a_i$ are fixed at $\mu_B=0$ (where $\sigma(\mu_B=0)=1)$ by fitting the DQPM coupling constant $g^2(T,{\mu_\mathrm{B} = 0})$ obtained from  Eq. (\ref{coupling_DQPM}) (see comparison of the basic thermodynamic observables from DQPM-CP and lQCD predictions in Fig. \ref{fig:eos_fix0}): $a_0=0.25$, $a_2=1.77$, $a_3=2.17$, $a_4=2.13$, $a_6=0.85$. 

The motivation to use the decrease in the effective coupling is based on the expectations of the QGP matter to approach the non-interacting Stefan-Boltzmann limit at large $\mu_B$ on the order of a few GeV and small temperatures $\mu_B \gg T$ (see recent pQCD results on the pressure in Ref. \cite{Kurkela:2016was}). Therefore, it is reasonable to assume that the coupling constant also decreases at $T_c$ with increasing $\mu_B$. To describe decrease of the coupling constant near the $T_c$ with $\mu_B$ we introduced an additional factor, affecting the region near the phase transition:
\begin{equation}
 \sigma(\mu_B)=1-\sigma_2\mu_B^2-\sigma_4\mu_B^4,   
\end{equation}
where $\sigma_2=0.45 GeV^{-2}$ and $\sigma_4=0.15 GeV^{-4}$. We fixed the values of $\sigma_2$ and $\sigma_4$ by fitting the quasiparticle entropy from Eq. (\ref{sdqp}) to the lQCD data points of the entropy density from the BMW collaboration \cite{Borsanyi:2012cr,Borsanyi:2013bia} at finite $\mu_B=0.1,0.2,0.3,0.4$ GeV for given temperature points $T_c(\mu_B)<T<T_{max}$, where $T_c(\mu_B)$ denotes the critical temperature and $T_{max}=0.4$ GeV. 

We note that the adjustment of the effective coupling constant is made in order to interpolate results for thermodynamic observables between the region of vanishing baryon chemical potential and that of the high baryon chemical potential. High and moderate temperatures above the phase transition line $T>T_c(\mu_B)$ are considered. The aim of the model is to describe on the one side qualitatively the behavior of the thermodynamic observables in the regions of high/moderate baryon density for $T>T_c(\mu_B)$, and on the other side to reproduce the lQCD EoS the region of moderate baryon chemical potential. For quantitative results one has to refer to  more rigorous approaches. 
To verify the $\mu_B$-dependence of the effective coupling we compare in Fig. \ref{fig:p_pqcd} the ratio $p/p_{SB}(T=0)$ ($p_{SB}(T=0) = \dfrac{\mu_B^4}{108 \pi^2}$) from DQPM-CP calculations for high $\mu_B$ with the pQCD calculations from  Ref. \cite{Kurkela:2016was}. Although the uncertainties of the pQCD results are quite large, we see that the resulting pressure from the DQPM-CP is compatible with the pQCD predictions.\\
\begin{figure}[h!]
\centering
\includegraphics[width=0.5\textwidth]{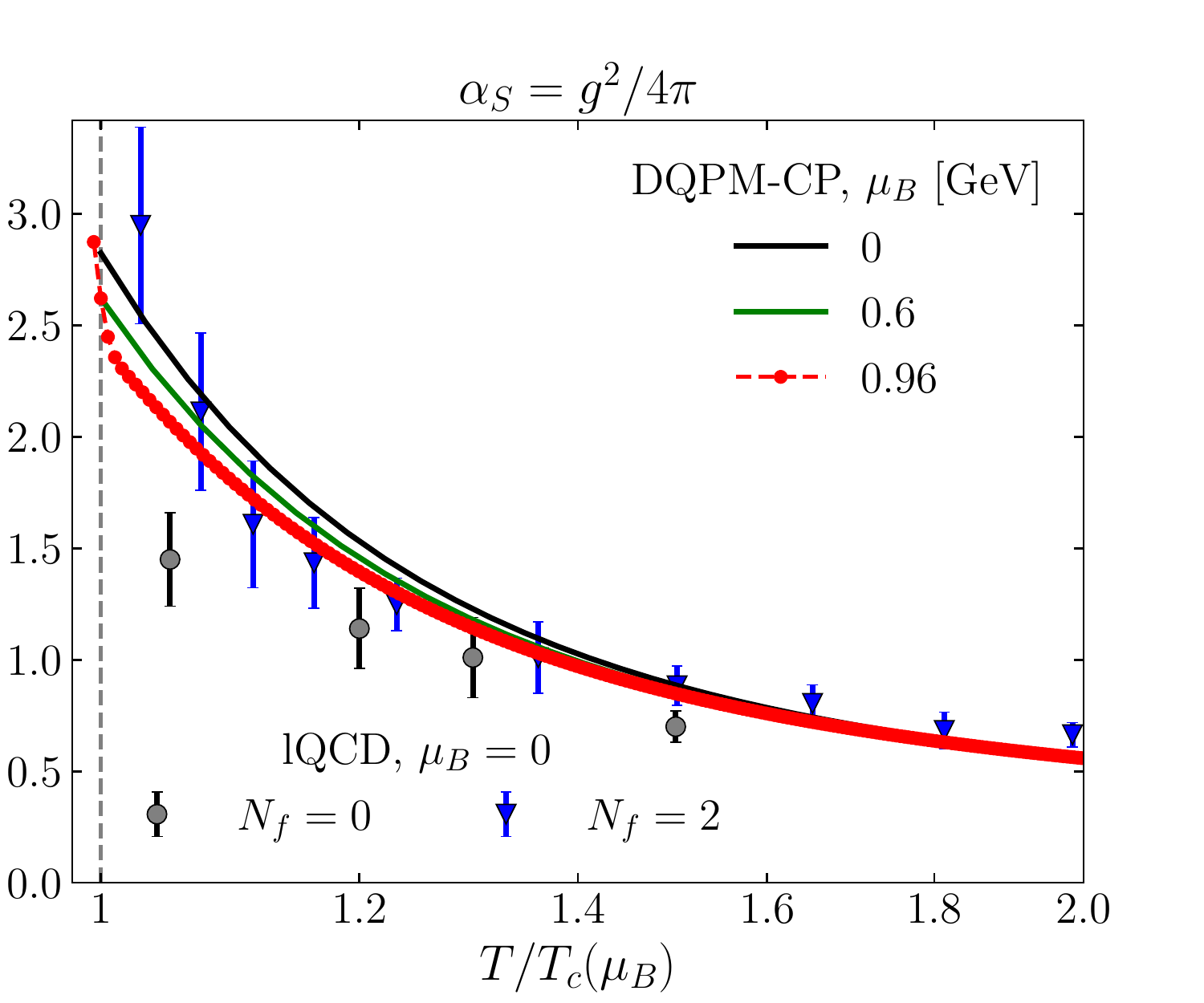}
\caption{\label{fig:as_fix0}  The running coupling $\alpha_S^{DQPM-CP}$ from Eq. (\ref{eq:as_DQPM-CP}) as a function of the scaled temperature $T/T_c$ at fixed $\mu_B$: for $\mu_\mathrm{B}=0$, $T_c=0.158$ GeV (black solid line) corresponds to the crossover phase transition and  ($T^{CEP}$,
$\mu^{CEP}_B) = (0.100, 0.960)$ GeV (red points) corresponds to the CEP. The lattice
results for quenched QCD, $N_f = 0$, (black circles) are taken from Ref. \cite{Kaczmarek:2004gv} and for $N_f = 2$ (blue triangles) are taken from Ref. \cite{Kaczmarek:2005PRD}.}
\label{fig:gs}
\end{figure}

Furthermore, to accumulate 'critical' behaviour near the CEP, where the phase transition is of second order, we use an additional 'critical' term for the coupling constant. The goal of this term is to describe the critical behaviour at the second order phase transition for the microscopic and thermodynamic quantities.
To obtain the parametrization of the 'critical' coupling constant, we fit the entropy density to the results from the PNJL \cite{Fuseau:2019zld}. 
The resulting parametrization for the 'critical' coupling constant is given by 
\begin{equation}
 \alpha_S^{crit}=a \cdot(T/T_c)^{-12}, 
 \label{eq:a_cep}
\end{equation}
where $a=\alpha_{S}^{cross}(T=T_{CEP})$.\\
The total coupling constant $\alpha_S^{DQPM-CP}$ then reads 
\begin{equation}
\alpha_S^{DQPM-CP} \equiv
 \left\{
\begin{aligned}
& \mu_B=\mu_{CEP}: \ \alpha_S^{CEP}  =    
\\&  \ \ \ \frac{1-F(T)}{2} \alpha_S^{crit} +  \frac{1+F(T)}{2} \alpha_{S}^{cross}  \\
& \mu_B \neq \mu_{CEP}: \ \alpha_{S}^{cross}
\end{aligned}
\right .
\label{eq:as_DQPM-CP}
\end{equation}
 $\alpha_{S}^{cross}$ corresponds to the coupling constant for the crossover region defined by Eq. (\ref{eq:a_cross}), while at $\mu_B=\mu_{CEP}=0.960$ GeV the effective coupling  $\alpha_S^{DQPM-CP}=\alpha_S^{CEP}$ includes the additional 'critical' contribution $\alpha_S^{crit}$, defined by Eq. (\ref{eq:a_cep}).
 To match the two coupling constants we employ the smoothing function:
 \begin{equation}
    F(T)= \mathrm{tanh} \left[\frac{T-0.1004}{\delta T}\right], 
 \end{equation}
 where $\delta T=0.002$ GeV is the region in the vicinity of the CEP, where the two coupling constants have to match. 
The values of $\delta T$ and $T$ are chosen in accordance with the $T/T_c$ - dependence of the PNJL entropy density. While the temperature $T_0=0.1004$ GeV regulates the size or temperature range $(T_{CEP}, T_0)$ of the critical contribution, $\delta T$ affects the derivatives of the pressure at $T_0$. However,  a slight change of $T_0$ as well as $\delta T$ up to 20\% will not change the qualitative results; the effect of an increase of the second-order derivatives will be less pronounced for smaller $T_0$.  

\begin{figure}
\centering
\begin{minipage}[h]{1\linewidth}
\center{\includegraphics[width=0.97\linewidth]{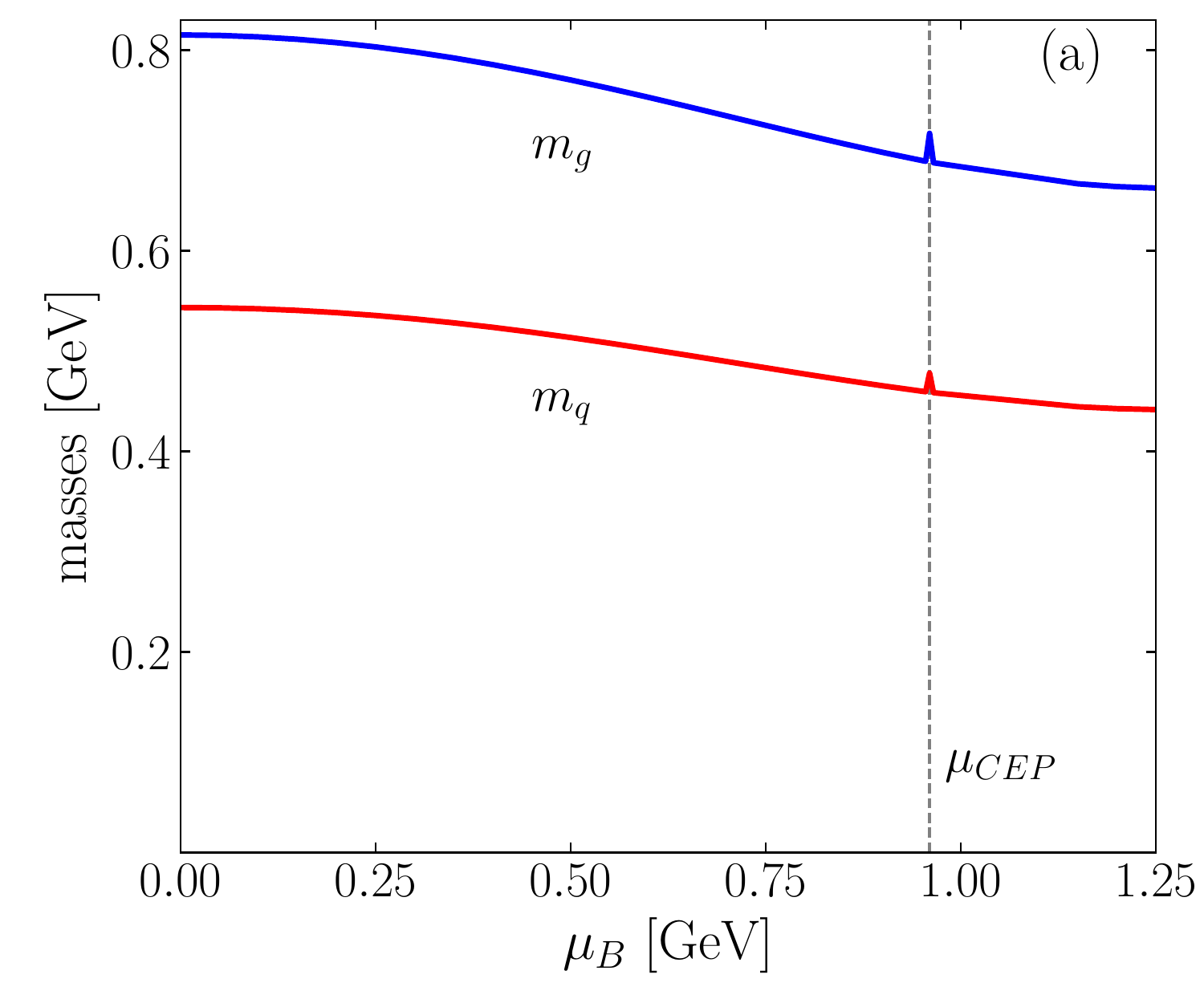}} 
\end{minipage}
\begin{minipage}[h]{1\linewidth}
\center{\includegraphics[width=0.97\linewidth]{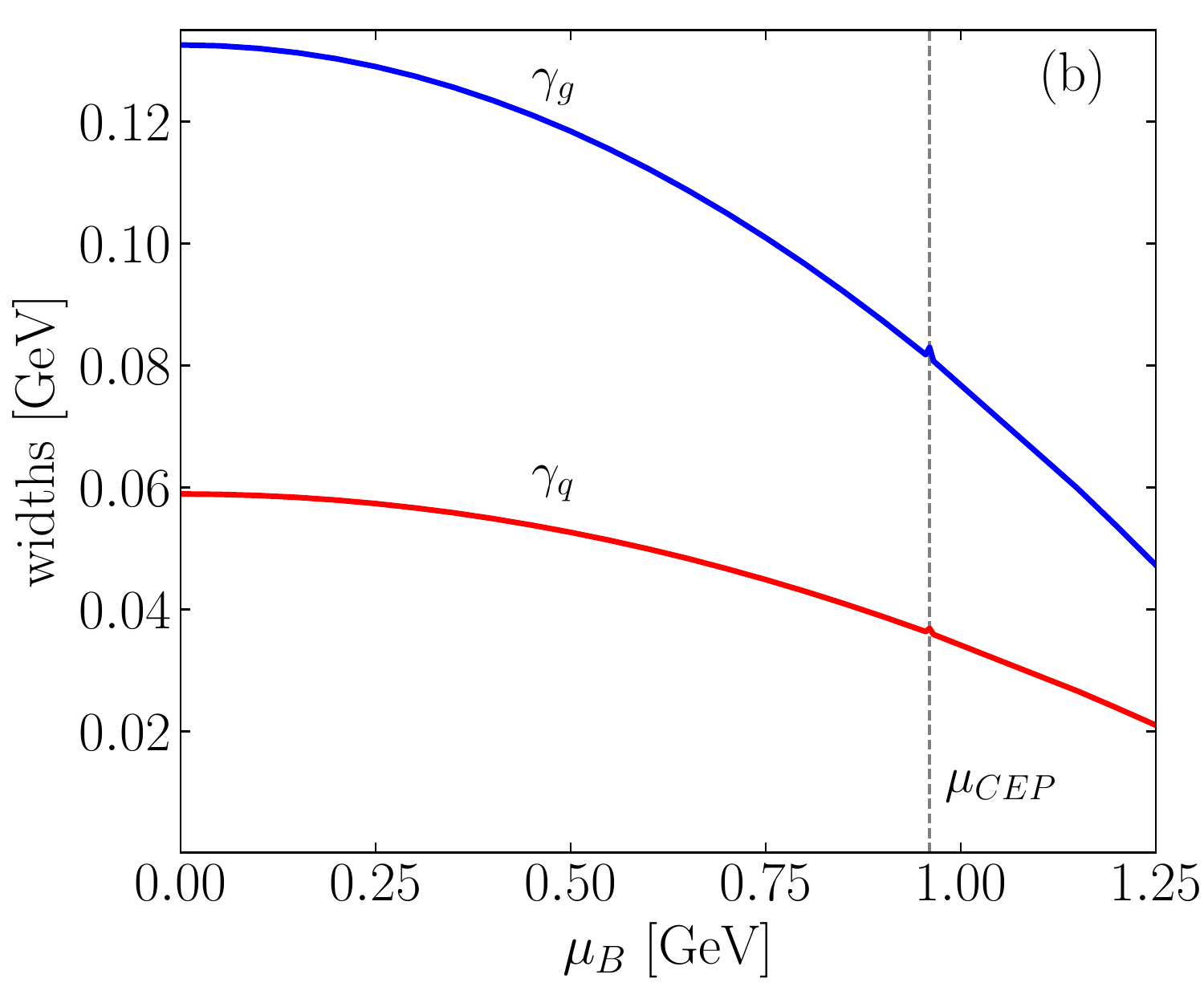}} 
\end{minipage}
\caption{\label{fig:mass_Tc} Effective masses (a) and widths (b) of light quarks and gluons in the DQPM-CP from Eqs. (\ref{polemass_q_dqpm}-\ref{eq:widths}) along the critical line (given by Eq. (\ref{eq:Tc_line})) as function of baryon chemical potential $\mu_B=3\mu_q$. The dashed line represents the critical value of the baryon chemical potential $\mu_{CEP}=0.96$ GeV.}
\end{figure}

 Figure \ref{fig:gs} shows the effective coupling of the DQPM-CP at fixed $\mu_B=0$ (black solid line) and $\mu_B=0.96$ GeV (red points) as a function of the scaled temperature $T/T_c$. At $\mu_B=0$ the coupling constant equals the DQPM effective coupling $g^2(T,{\mu_\mathrm{B} = 0})$ obtained from  Eq. (\ref{coupling_DQPM}). \\
The scaling behavior of the quasiparticle masses has been employed in condensed matter physics \cite{Wolfle:2016utb}, where the interaction with bosonic fluctuations near the critical point causes a divergence in the effective masses of the quasiparticles.
We note that one can include the scaling behavior of the thermodynamic observables in a more rigorous way as done in Ref. \cite{Parotto:2018pwx}, where the EoS from the lQCD calculations of the BMW collaboration has been parametrized and adopted to include a singular part near the CEP from the 3D-Ising model. 

The effective masses (a) and widths (b) of quarks and gluons along the critical line are depicted on Fig. \ref{fig:mass_Tc}. The masses decrease with increasing of $\mu_B$, while at the CEP they show a peak, which corresponds to the finite value of the 'critical' coupling constant. Since the effective masses reflect medium modifications, the increase when approaching the CEP is expected. The thermal widths should accordingly manifest a similar behavior. 
The widths of the quarks and gluons are significantly smaller then the masses for the whole $\mu_B$-range. 
The ratio of the pole mass to the width follows from Eqs.~(\ref{polemass_g_dqpm}-\ref{eq:widths}):
\begin{equation}
    m_i/\gamma_i \propto \dfrac{a_i + b_i \dfrac{\mu^{2}_{q}}{T^2\pi^2}} {g(T,\mu_B)\ln\left(2c_m/g^2(T,\mu_\mathrm{B})+1\right)},
\end{equation}
where we use the shorthand notation for constants $a_i = 1$ (for quarks), $1+\frac{N_{f}}{2N_{c}}$ (for gluons), $b_i=1$ (for quarks),  3/2 (for gluons).
For vanishing chemical potential the ratio $m_i/\gamma_i \approx 9$ for quarks and $\approx 6$ for gluons. The ratio increases with $\mu_B$ since the coupling constant decreases with $\mu_B$, for instance at $\mu_B=1$ GeV: we set $m_i/\gamma_i \approx 16$ for quarks and $\approx 11$ for gluons.

Importantly, we see that in DQPM-CP the quark masses are larger than a third of the free proton mass. This means that the production of baryons across the critical line, the dominant process for large $\mu_B$ and small $T$, is an exothermic process in DQPM-CP.

\begin{figure}[h!]
\centering
\includegraphics[width=0.52\textwidth]{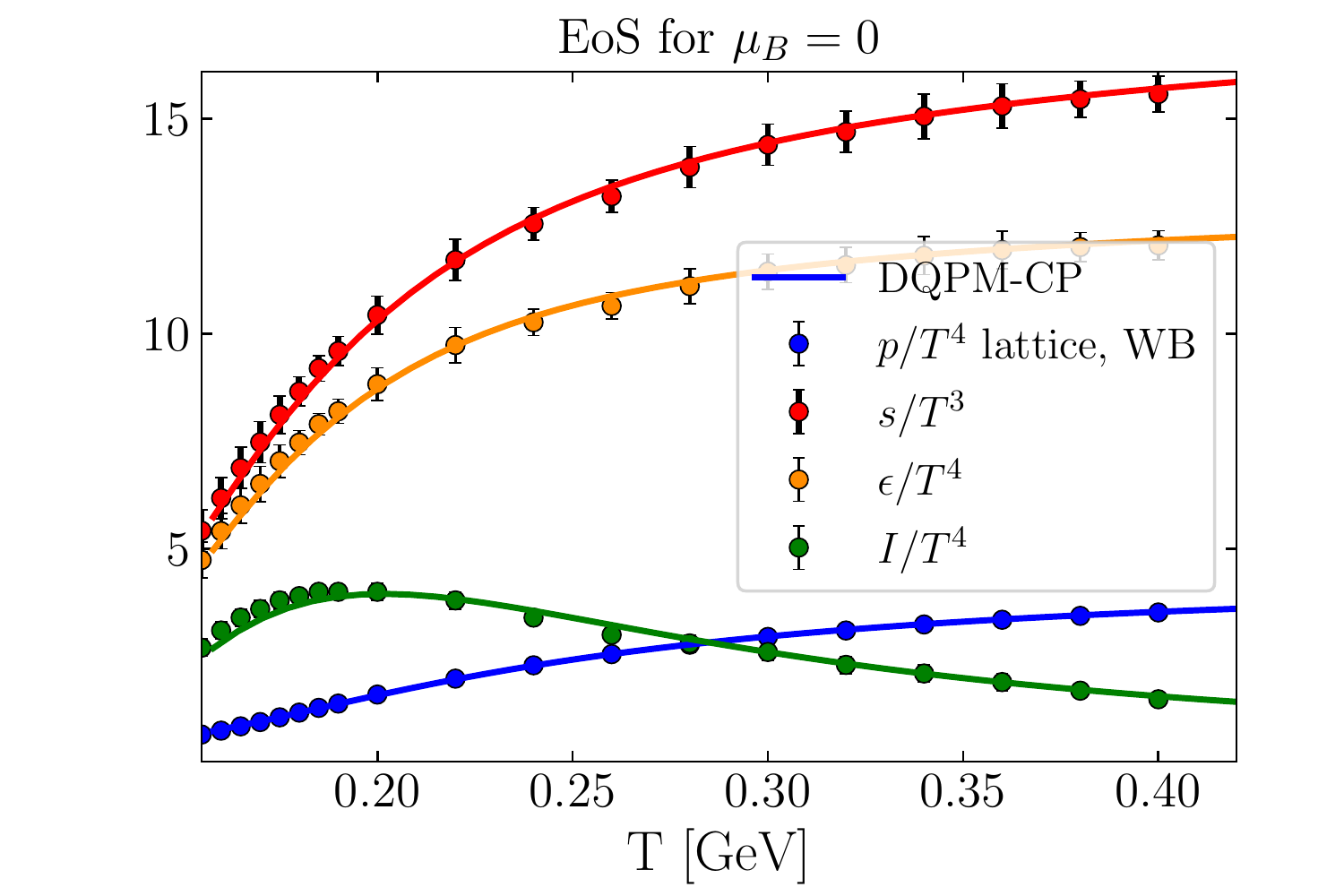}
\caption{\label{fig:eos_fix0} The scaled pressure $P(T)/T^4$ (blue line), entropy density $s(T)/T^3$ (red line), scaled energy density $\epsilon(T)/T^4$ (orange line), and interaction measure $I(T)/T^4 $ (green line), from the DQPM-CP in comparison to the lQCD results from Refs. \cite{Borsanyi:2012cr,Borsanyi:2013bia} (circles) for zero baryon chemical potential.}
\end{figure}
~\\
\section{\label{sec3}EoS for finite temperature and chemical potential}

In this Section we consider the basic thermodynamic observables from the DQPM-CP for finite chemical potential.
Starting point for the calculation of the thermodynamic functions in the dynamical quasiparticle models is the evaluation of the entropy density and the quark densities  via the propagators as described in Eqs. (\ref{sdqp}) and (\ref{nbdqp}). 
\begin{figure}[!h]
 \centering
\begin{minipage}[h]{0.905\linewidth}
\center{\includegraphics[width=1.0\linewidth]{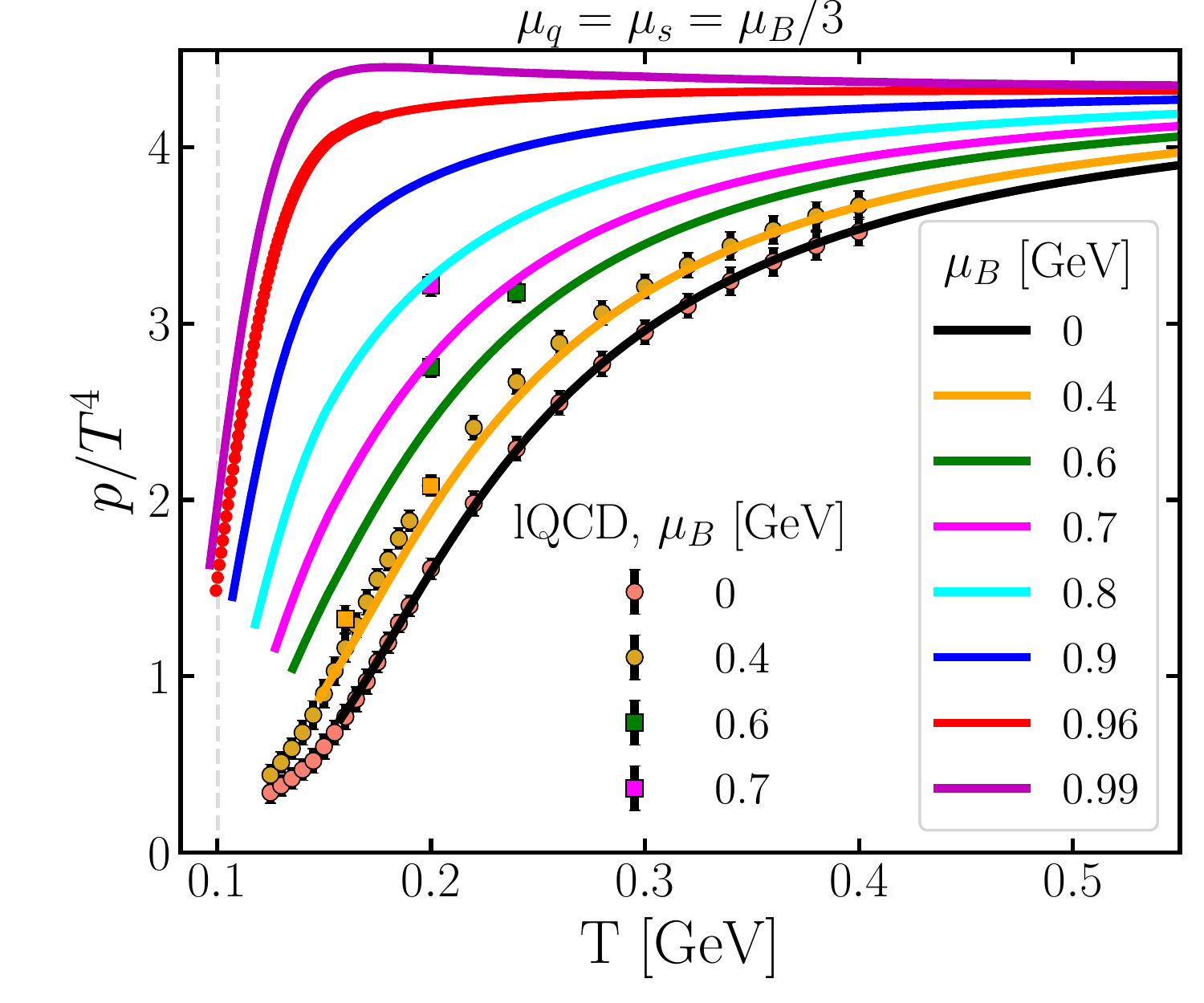} }
\end{minipage}
\begin{minipage}[h]{0.905\linewidth}
\center{\includegraphics[width=1.0\linewidth]{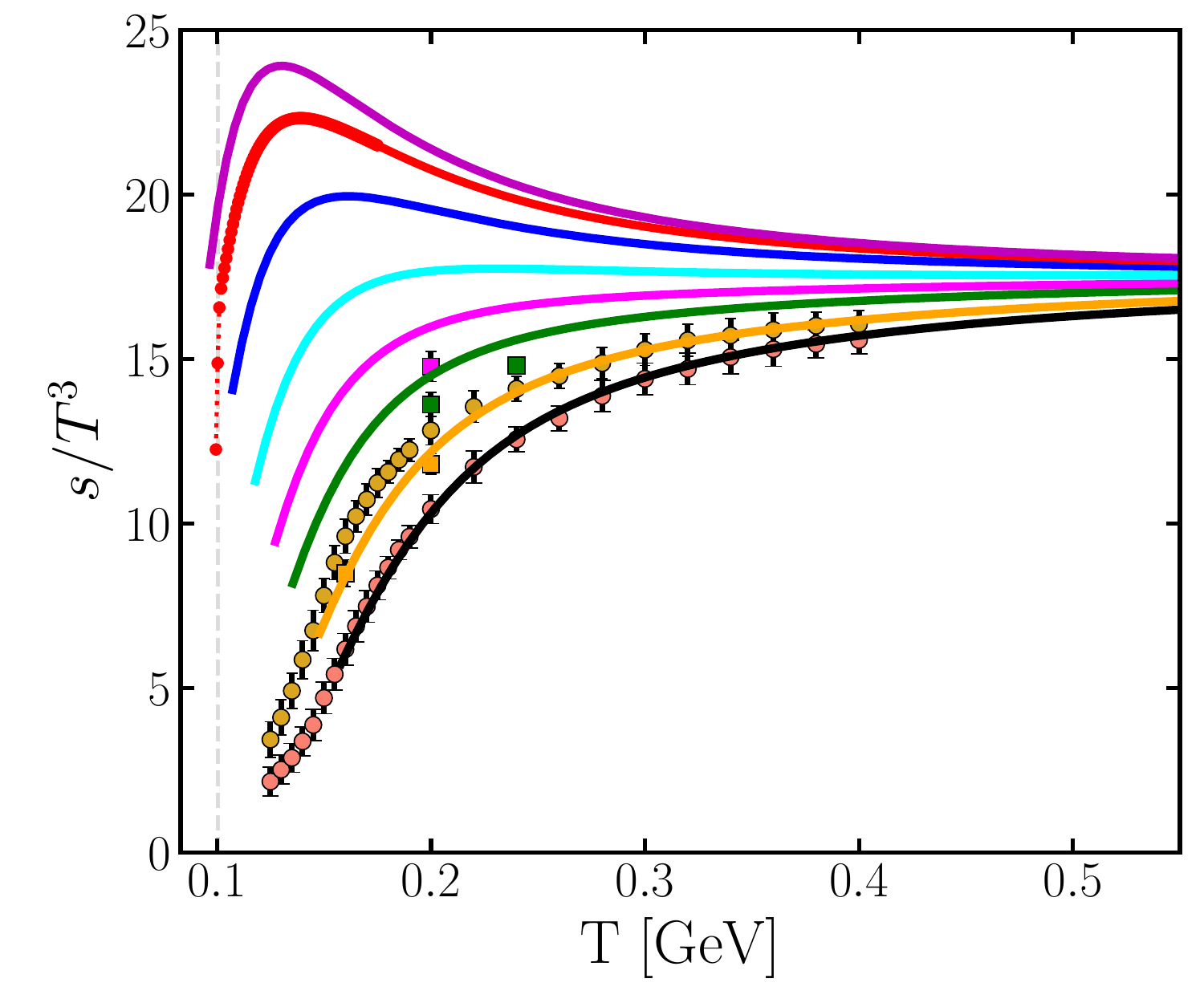}}
\end{minipage}
\begin{minipage}[h]{0.905\linewidth}
\center{\includegraphics[width=1.0\linewidth]{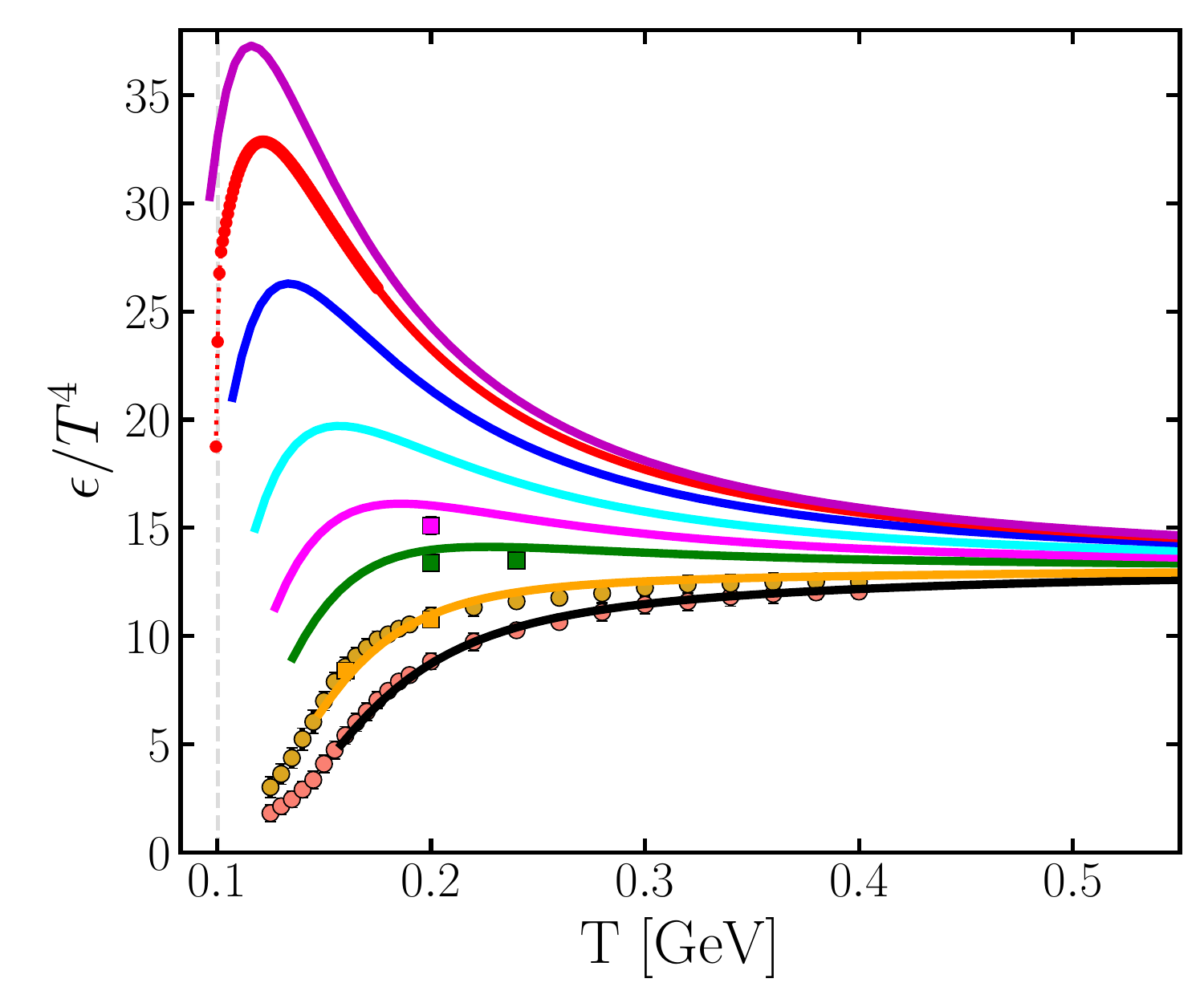}}
\end{minipage}
  \caption{Scenario: $\mu_q=\mu_u=\mu_s=\mu_B/3$.
  From top to bottom: Scaled pressure $p/T^4$, entropy density $s/T^3$, and scaled energy density $\epsilon/T^4$  from the DQPM (lines) as a function of temperature $T$ at various values of $\mu_\mathrm{B}$ [GeV]. The lQCD results obtained by the BMW group are taken from Refs. \cite{Borsanyi:2012cr,Borsanyi:2013bia} (circles) and from Ref. \cite{Borsanyi:2021sxv} (squares). The dashed line displays the critical temperature $T_{CEP}=0.10$ GeV. }
  \label{fig:eos_2d_mub}
    \end{figure}
    ~\\
        ~\\
                ~\\
\begin{figure}[!h]
 \centering
\begin{minipage}[h]{0.905\linewidth}
\center{\includegraphics[width=1.0\linewidth]{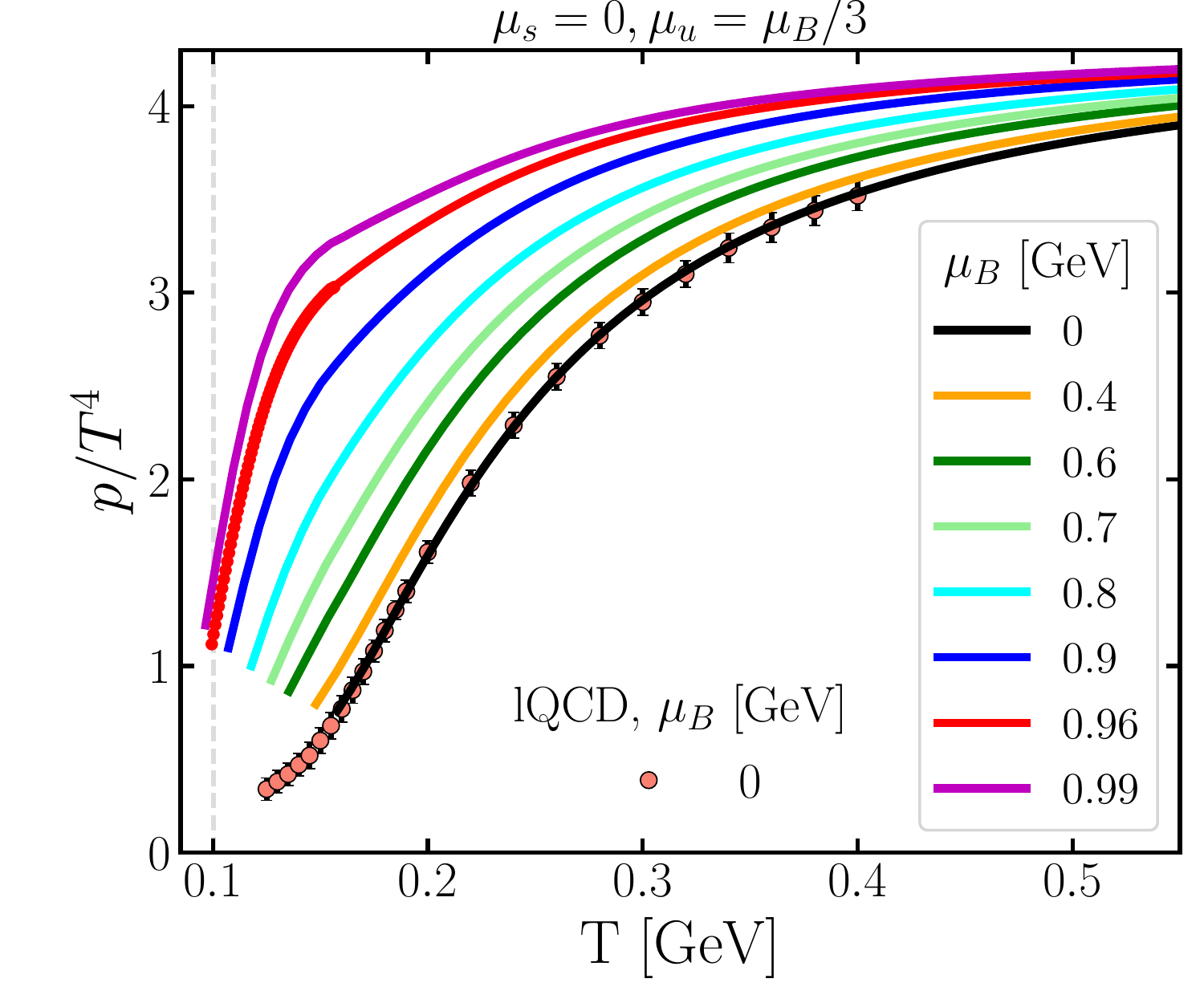} }
\end{minipage}
\begin{minipage}[h]{0.905\linewidth}
\center{\includegraphics[width=1.0\linewidth]{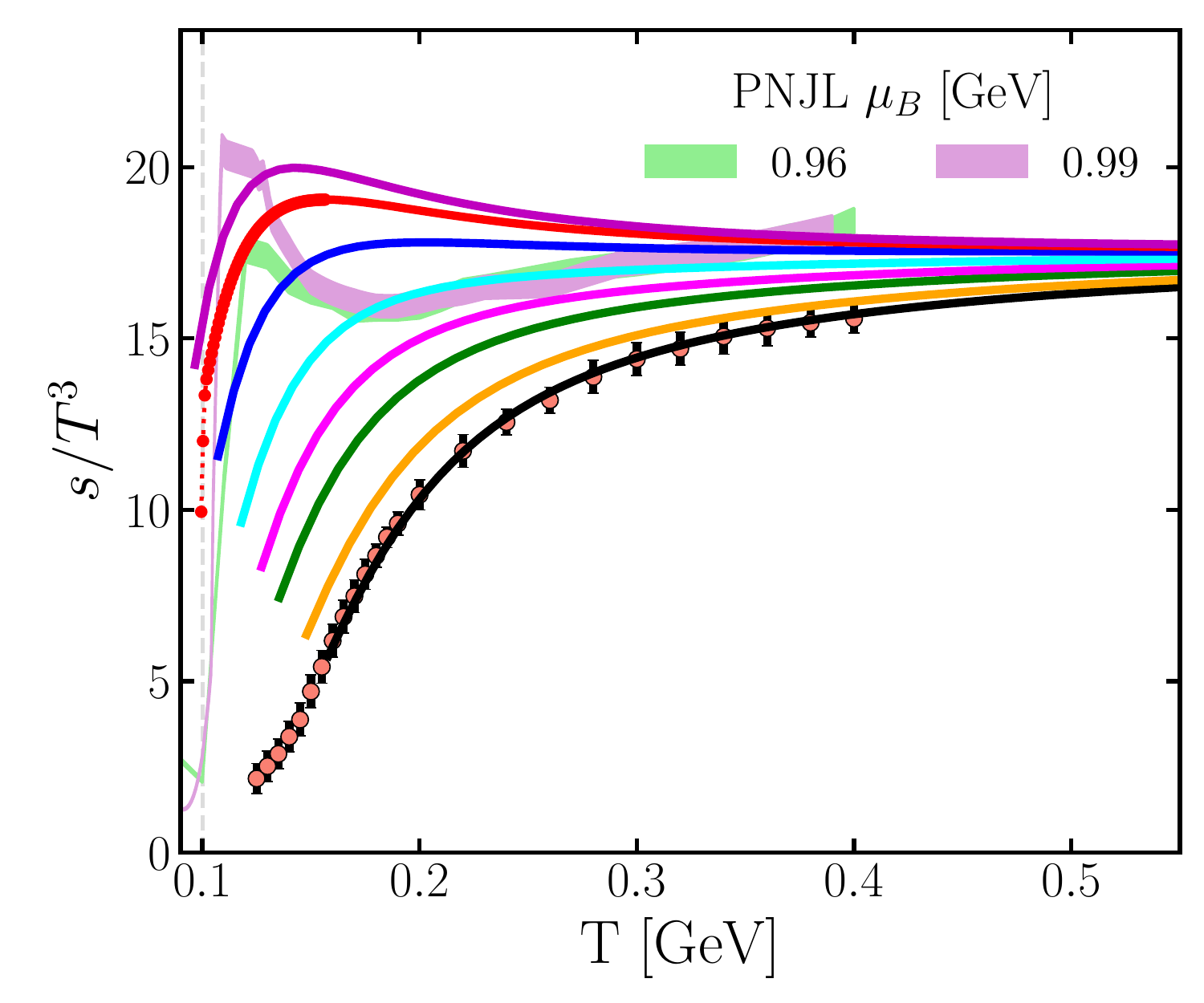}}
\end{minipage}
\begin{minipage}[h]{0.905\linewidth}
\center{\includegraphics[width=1.0\linewidth]{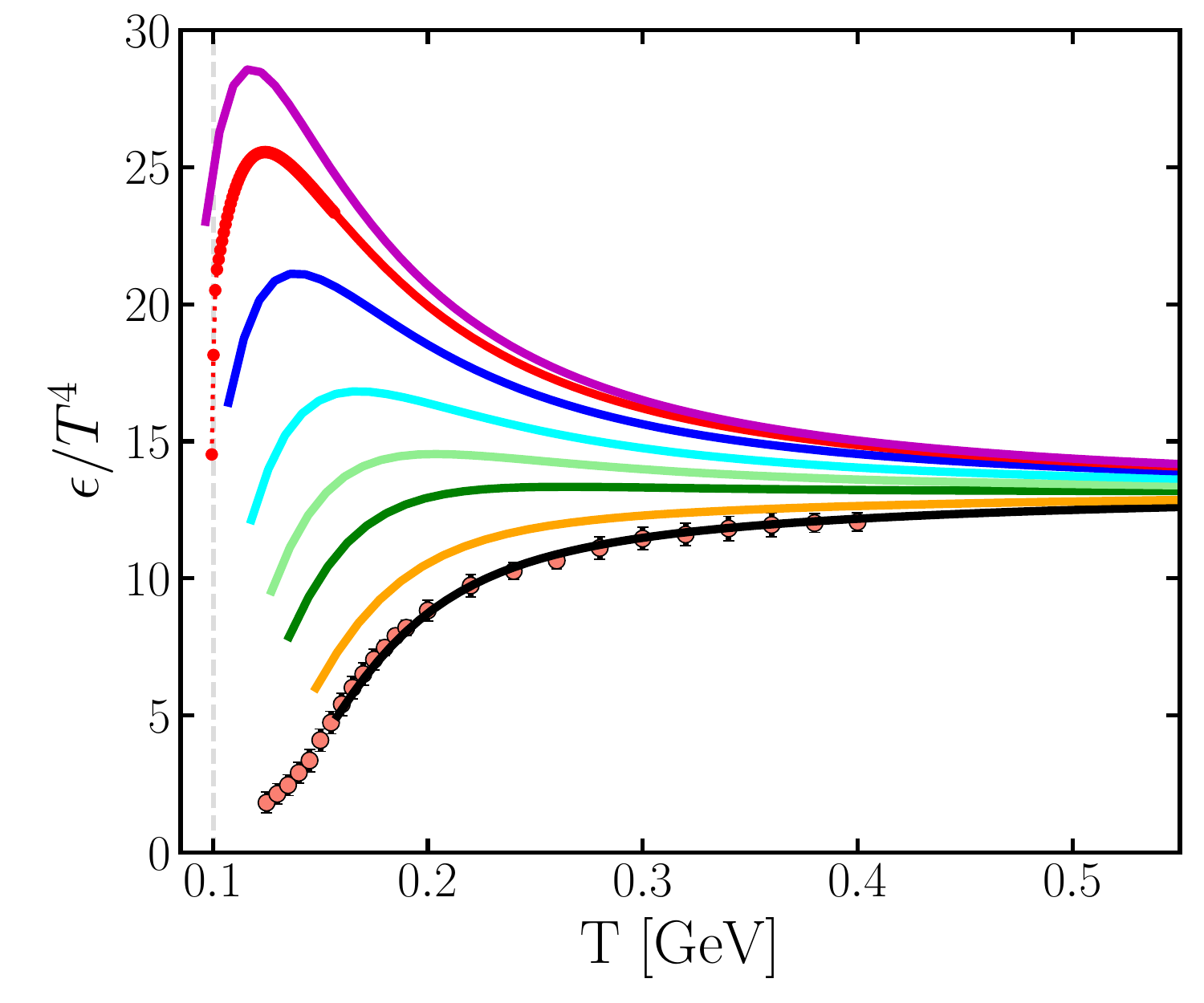}}
\end{minipage}
  \caption{Scenario: $\mu_s=0,\mu_u=\mu_B/3$.
  From top to bottom: Scaled pressure $p/T^4$, entropy density $s/T^3$, and scaled energy density $\epsilon/T^4$  from the DQPM-CP (lines) as a function of temperature $T$ at various values of $\mu_\mathrm{B}$ [GeV]. The lQCD results obtained by the BMW group are taken from Refs. \cite{Borsanyi:2012cr,Borsanyi:2013bia} (circles). The PNJL results for the entropy density (colored area) are taken from Ref. \cite{Fuseau:2019zld}.}
  \label{fig:mus_eos_2d_mub}
    \end{figure}
Then it is straightforward to derive the pressure $p$ and later the energy density, employing the Maxwell relation 
for a grand canonical ensemble:	

		\begin{align}
	p(T,\mu_\mathrm{B}) =  p_0(T,0) + \int\limits_{0}^{\mu_\mathrm{B}} n_B(T,\mu_\mathrm{B}')\ d\mu_\mathrm{B}'. \label{pressure} 
	\end{align}

For the pressure at $\mu_B=0$ we use the lQCD parametrization of the pressure $p_0(T,0)$ from Ref.\cite{Borsanyi:2012cr,Borsanyi:2013bia}.
	The energy density $\epsilon$ then follows from the Euler relation
	\begin{equation}
	\label{eps} \epsilon = T s - p + \sum_i \mu_i n_\mathrm{i}.
	\end{equation}
Furthermore, the interaction measure is defined as:	
	\begin{equation}
	\label{wint} I \equiv \epsilon - 3P = Ts - 4p + \sum_i \mu_i n_\mathrm{i},
	\end{equation}
which vanishes in the non-interacting limit of massless degrees of freedom at $\mu_\mathrm{B} = 0$.
The scaled pressure, entropy density and energy density of the QGP phase are supposed to increase with the temperature. However, lQCD calculations of the thermodynamic observables show \cite{Weber:2018bam}, that the massless non-interacting limit can not be reached even at temperatures of $T\sim 1$ GeV. 

We consider two setups for the quark chemical potentials:
	(I) $\mu_q=\mu_u=\mu_s=\mu_B/3$ and (II) $\mu_s=0,\mu_u=\mu_B/3$.
The quark chemical potential can be related to the strange, baryon and electric charge chemical potentials as $\mu_{i}=B_{i}\mu_\mathrm{B} + Q_{i}\mu_\mathrm{Q} + S_{i}\mu_\mathrm{S}$ , where $B_{i}$,$Q_{i}$ and $S_{i}$ are baryon number, electric charge and strangeness of the considered quark. Herein we fix $\mu_Q=0$, therefore for the symmetric QGP matter (I) $\mu_q=\mu_u=\mu_s=\mu_B/3$ the strange and the electric charge  potentials are vanishing $\mu_S=\mu_Q=0$, while for (II) $\mu_s=0,\mu_u=\mu_B/3$ the strange chemical potential is finite $\mu_S=\mu_B/3$.

The T-dependence of the thermodynamic quantities such as the scaled entropy density, the pressure and the energy density from the DQPM-CP for various baryon chemical potentials $0\leq\mu_B\leq0.99$ GeV is shown in Fig. \ref{fig:eos_2d_mub} ($\mu_q=\mu_u=\mu_s=\mu_B/3$) and Fig. \ref{fig:mus_eos_2d_mub} ($\mu_s=0,\mu_u=\mu_B/3$). 
For setup (I) we found a good agreement between the DQPM-CP results (lines) and results from lQCD, obtained by the BMW group \cite{Borsanyi:2012cr,Borsanyi:2013bia} at $\mu_\mathrm{B} = 0$  and $\mu_\mathrm{B} = 400$ MeV.
The thermodynamical quantities increase with $\mu_B$. When
approaching the CEP at $\mu_B=0.96$ GeV the values of the entropy density, of the energy density as well as of the quark or baryon density rise suddenly.

For setup (II) we compare the results for the entropy density to that of the Nantes PNJL approach \cite{Fuseau:2019zld}.
The DQPM-CP results are in agreement with the PNJL results in the high temperature region $T\geq0.3$ GeV, while in the vicinity of the phase transition there is a clear deviation from the PNJL results, which can be expected since the two models encompass different microscopic properties of the degrees of freedom.
The resulting values of the thermodynamic observables for setup (II) is smaller than for setup (I) since the contribution from the strange quarks to the quasiparticle entropy density (see Eq. (\ref{sdqp})) is smaller for $\mu_s=0$ mainly due to the derivatives $\dfrac{\partial f_q(\omega-\mu_q)}{\partial T}$.
\begin{figure}[!ht]
\centering
\includegraphics[width=0.52\textwidth]{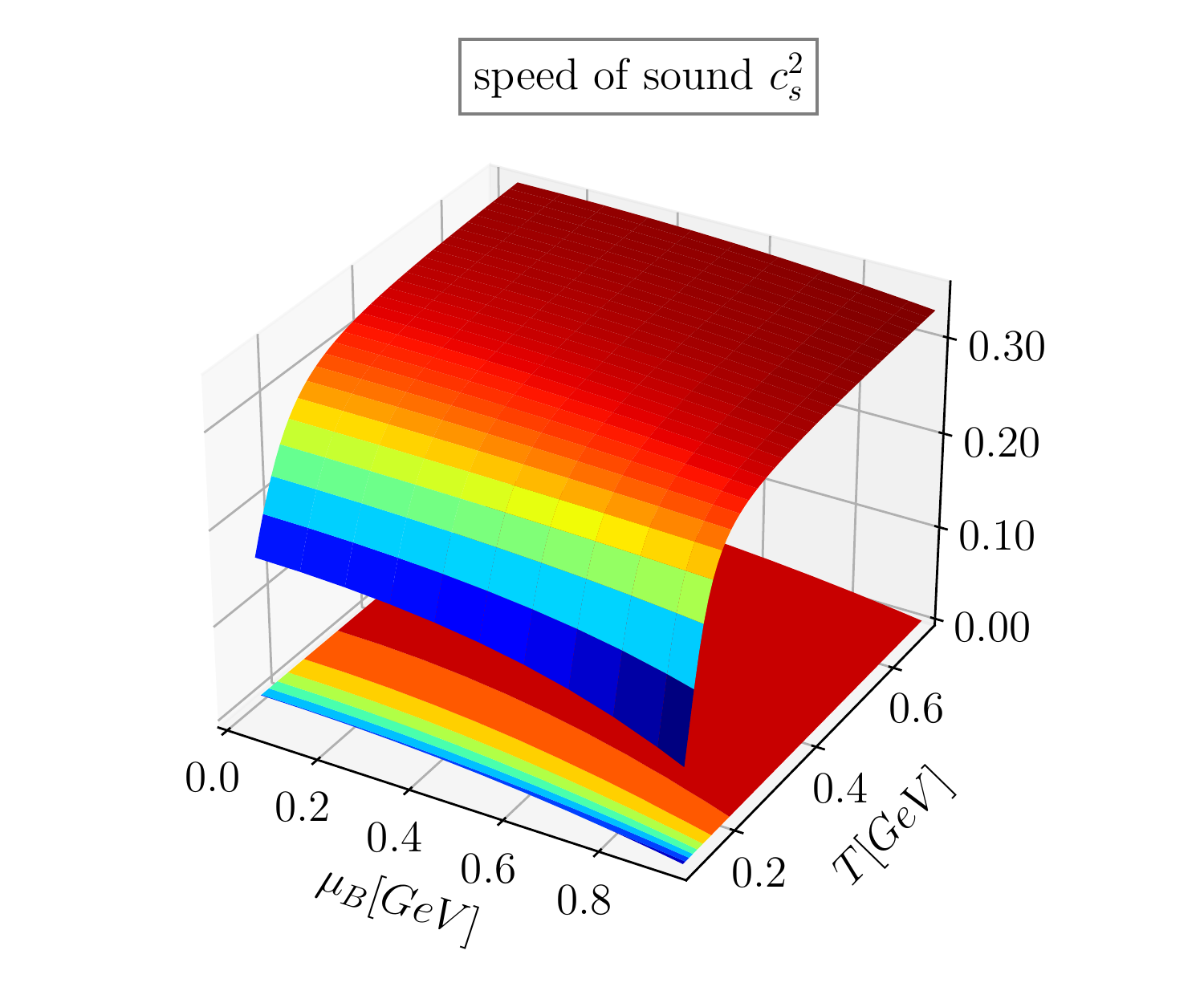}
\caption{\label{fig:3d_cs_muB} Scenario: $\mu_q=\mu_u=\mu_s=\mu_B/3$.
The speed of sound squared $c_s^2$ from the DQPM-CP for a crossover phase transition ($0\leq\mu_B<0.96$) as a function of $T$ and $\mu_B$. }
\end{figure}
\begin{figure}[!ht]

\begin{minipage}[h]{1\linewidth}
\center{\includegraphics[width=0.92\linewidth]{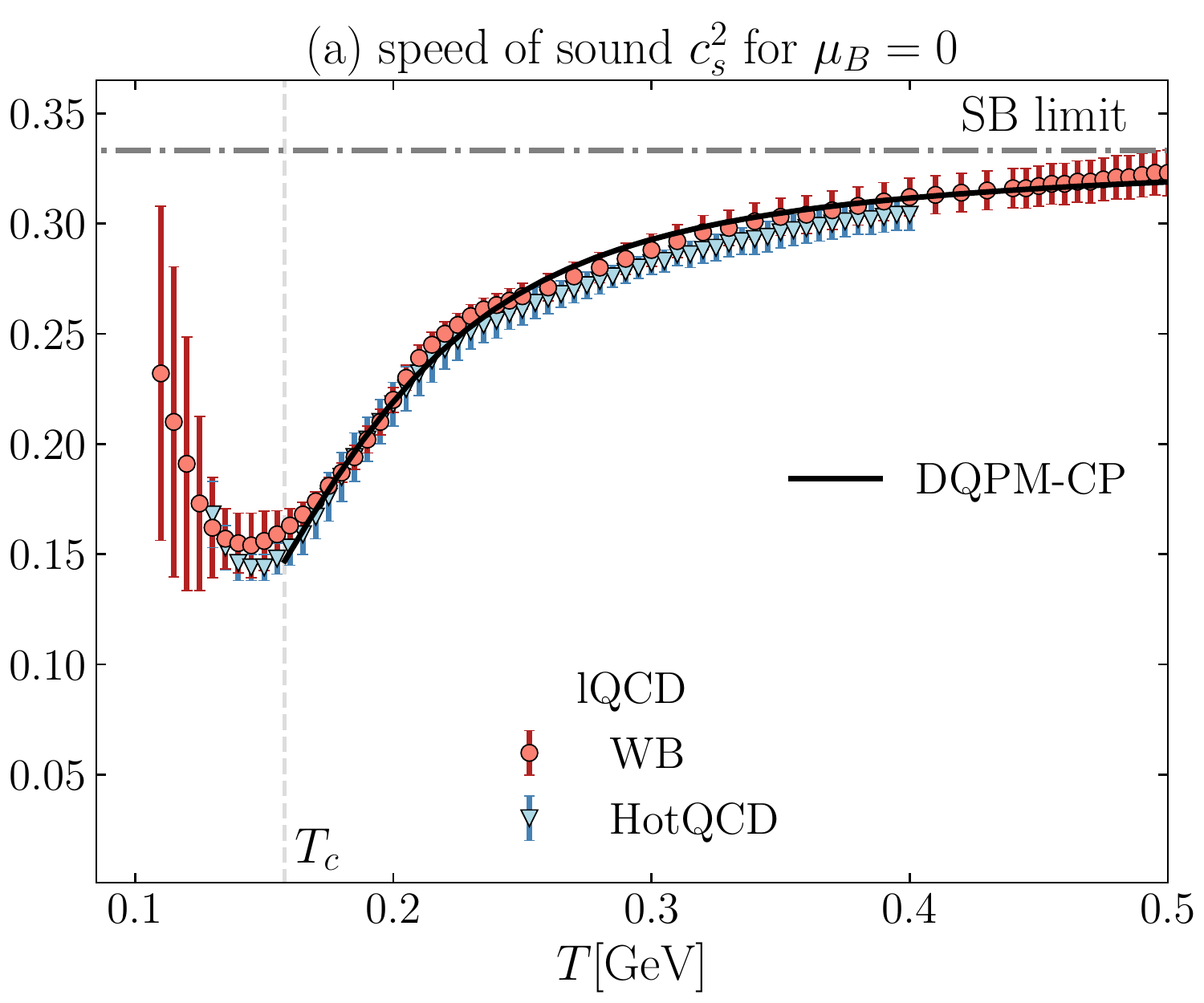}}
\end{minipage}
\begin{minipage}[h]{1\linewidth}
\center{\includegraphics[width=0.92\linewidth]{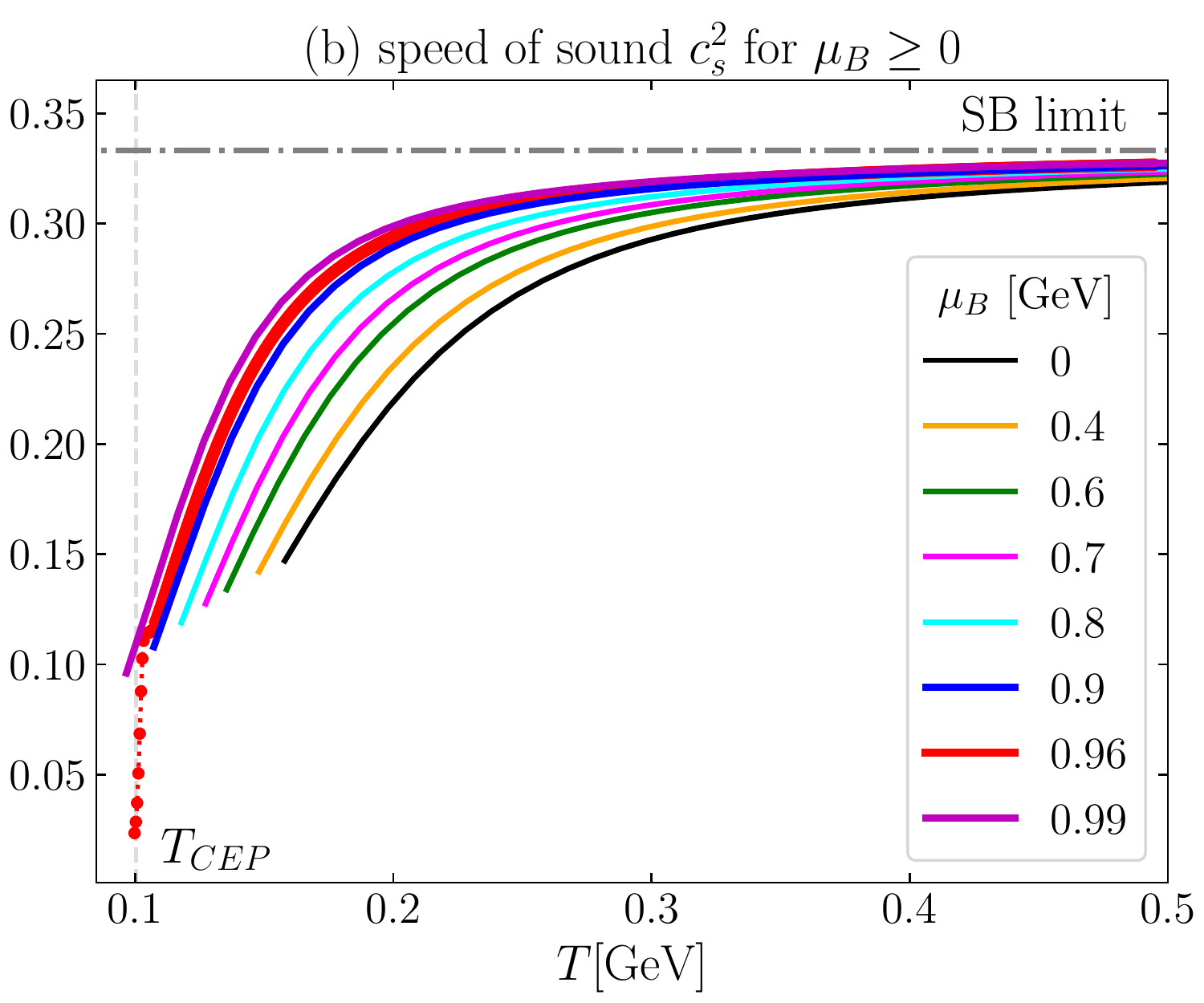}}
\end{minipage}
\caption{\label{fig:2d_cs_muB}Scenario: $\mu_q=\mu_u=\mu_s=\mu_B/3$.
The speed of sound squared $c_s^2$ from the DQPM-CP for (a) $\mu_B=0$ and (b) $\mu_B\geq0$ as a function of $T$ compared to lQCD results for $\mu_B=0$ obtained by the Wuppertal-Budapest collaboration \cite{Borsanyi:2013bia}(light red circles) and the HotQCD collaboration  \cite{HotQCD:2014kol} (blue triangles).  }
\end{figure}
\begin{figure}[!ht]
\centering
\includegraphics[width=0.5\textwidth]{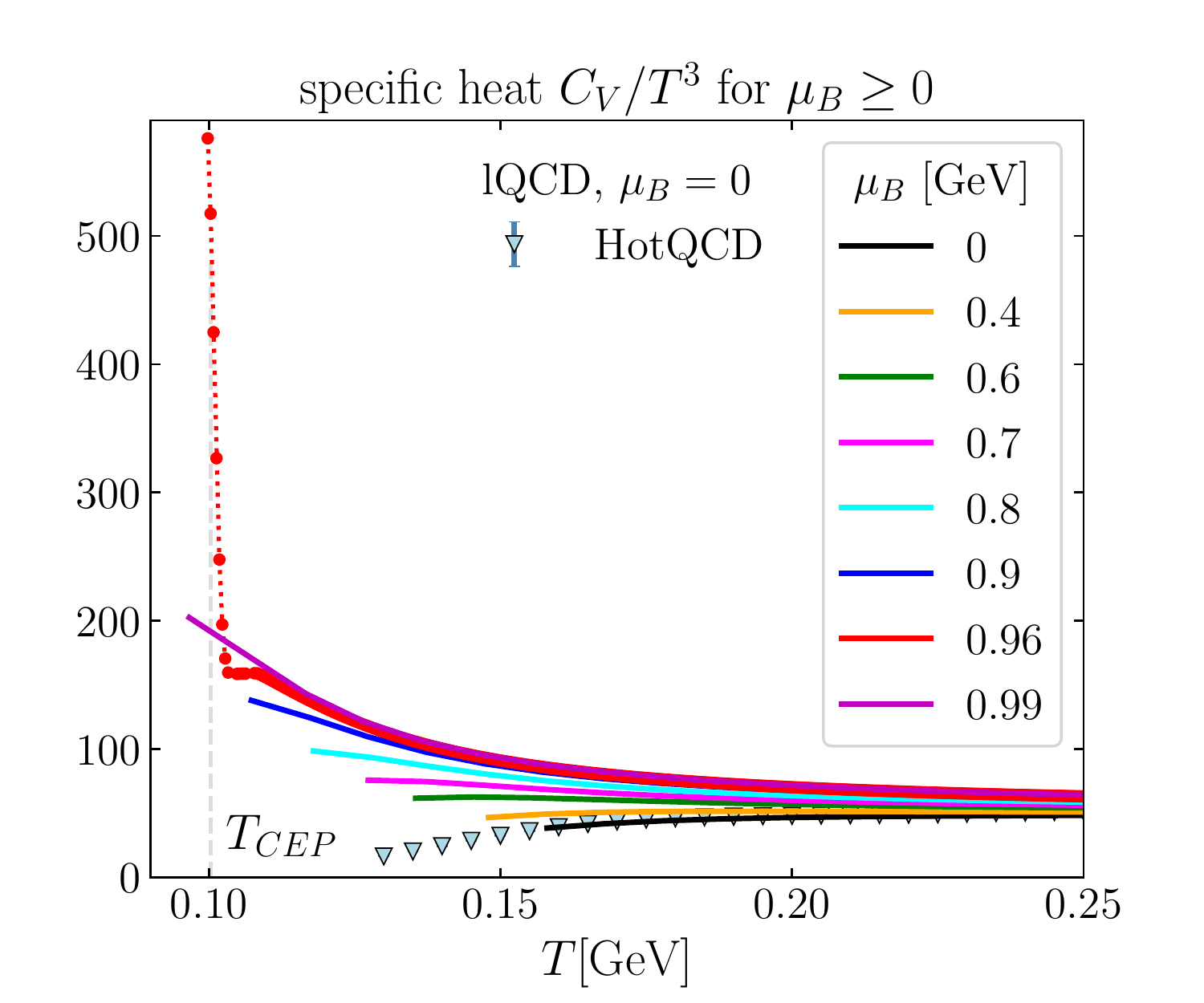}
\caption{\label{fig:2d_cv_muB}Scenario: $\mu_q=\mu_u=\mu_s=\mu_B/3$.
The spefic heat $C_V/T^3$ from the DQPM-CP at fixed $\mu_B$ as a function of $T$ compared to lQCD results for $\mu_B=0$ of the HotQCD collaboration  \cite{HotQCD:2014kol}. }
\end{figure}

\subsection{Approaching the CEP from the deconfined phase}

To realize a critical behaviour of the thermodynamic observables in the vicinity of the CEP we  introduce, as described in Sec. \ref{sec2} B, the 'critical' contribution to the coupling constant that affects the microscopic and macroscopic quantities. 
At the CEP, where the transition is of second order, the entropy density and baryon density increase rapidly but remain finite, while the quark susceptibility and the specific heat $C_V/T^3= \dfrac{d \epsilon}{d T} $ diverge. Therefore, the speed of sound (squared) vanishes as one approaches the CEP.
We consider the speed of sound and the specific heat at fixed $\mu_B$.
For fixed $\mu_B$ the speed of sound can be expressed as:
\begin{align}
    c_s^2=\frac{dp}{d\epsilon}=
  \frac{dp/dT}{d\epsilon/dT}=\frac{s}{C_V}. 
  \label{eq:cs2}
\end{align}
\begin{figure}
\centering
\begin{minipage}[h]{0.94\linewidth}
\center{\includegraphics[width=1.\linewidth]{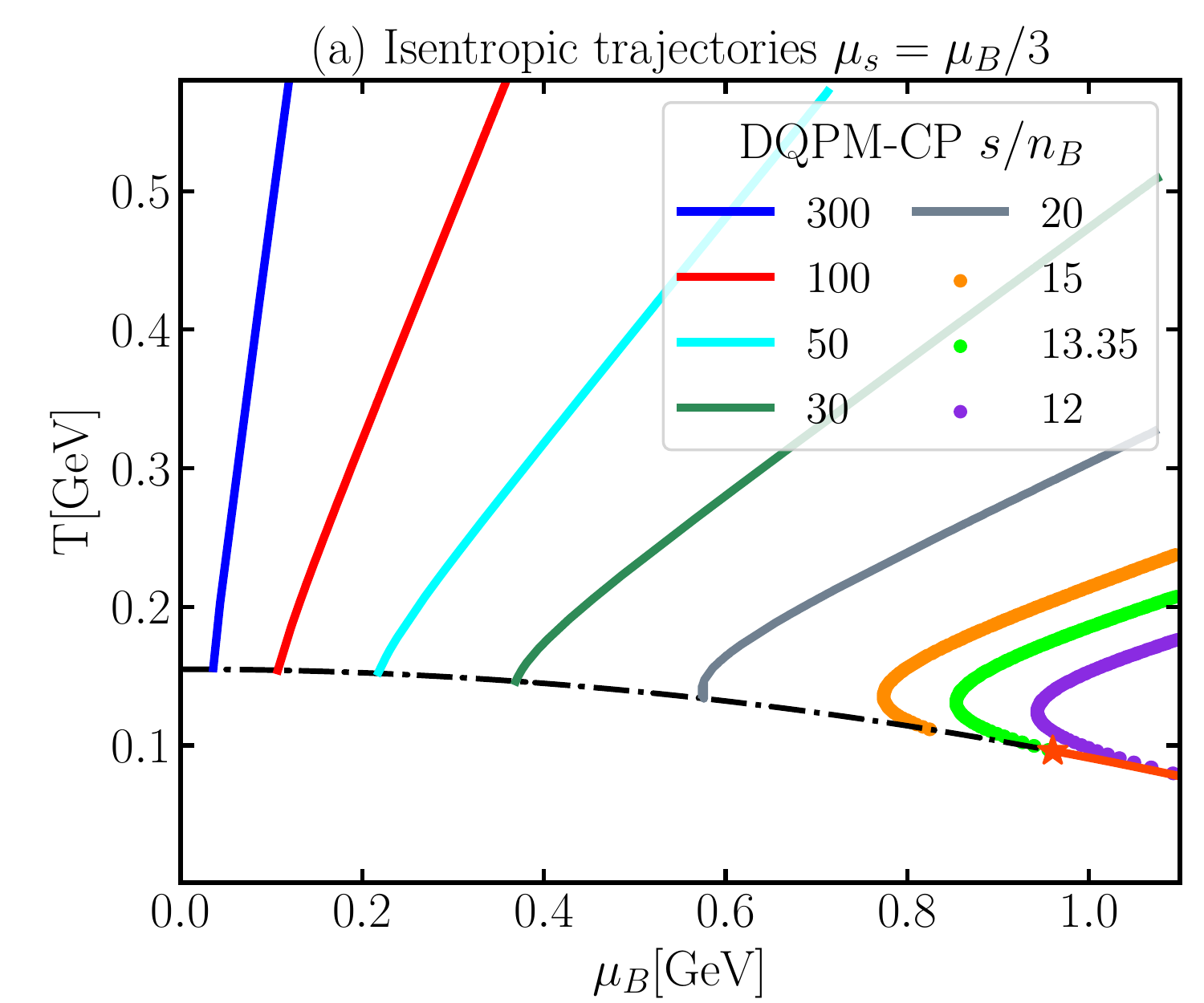} }
\end{minipage}
\begin{minipage}[h]{0.94\linewidth}
\center{\includegraphics[width=1.\linewidth]{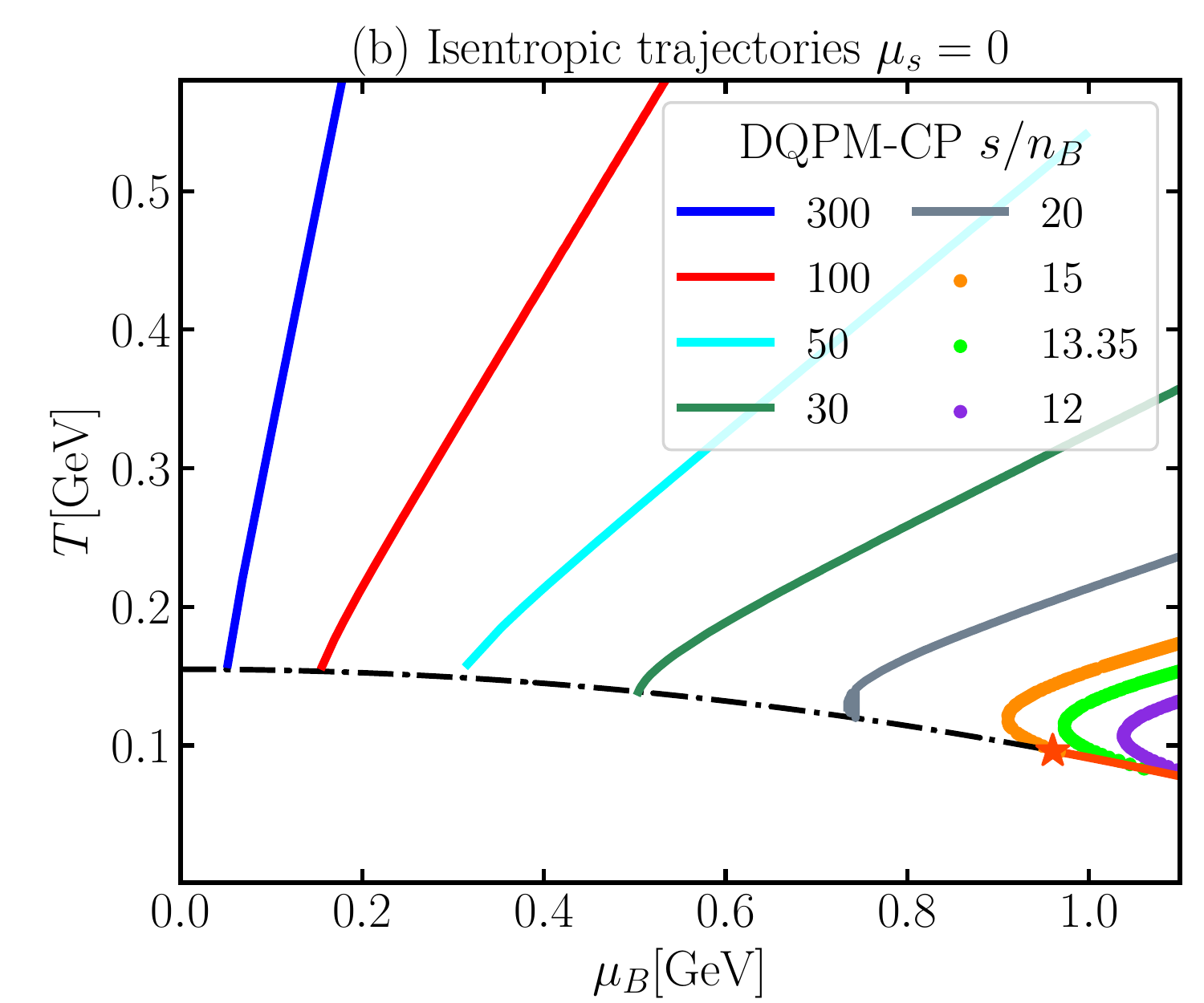}}
\end{minipage}
\begin{minipage}[h]{0.94\linewidth}
\center{\includegraphics[width=1.\linewidth]{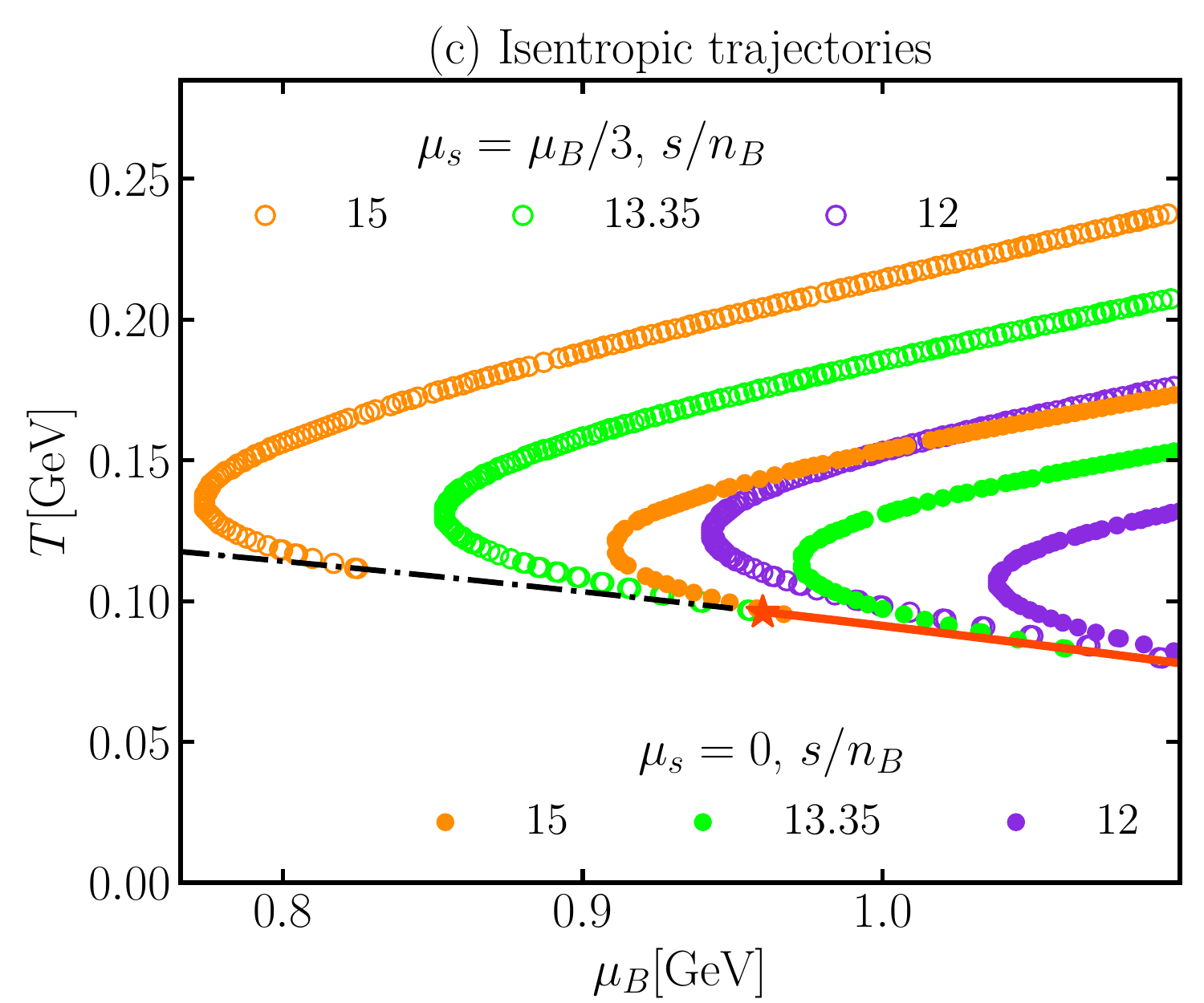}}
\end{minipage}

\caption{\label{fig:isentropes}Trajectories of constant $s/n_B$ in the DQPM-CP phase diagram for $T>T_c$ and (a) $\mu_q=\mu_u=\mu_s=\mu_B/3$, (b) $\mu_s=0,\mu_u=\mu_B/3$ and (c) two cases in the vicinity of the critical endpoint CEP. The finite temperature crossover (black dash-dotted line) at small chemical potential switches to the large chemical potential first-order transition (red solid line) at the CEP (star), which is located at $(0.10, 0.96)$ GeV.}
\end{figure}

The speed of sound squared in the DQPM-CP is depicted in Fig. \ref{fig:3d_cs_muB} as a function of temperature $T$ and baryon chemical potential $\mu_B$ in the crossover region, where $\mu_B \leq$ 0.95 GeV. The resulting $c_s^2$ increases with temperature and decreases near the phase transition with increasing $\mu_B$. 
In Fig. \ref{fig:2d_cs_muB} (a) we show the comparison of the DQPM-CP results for $c_s^2$ at vanishing $\mu_B$ with the available lQCD estimations from the Wuppertal-Budapest collaboration \cite{Borsanyi:2013bia} (light red circles) and the HotQCD collaboration  \cite{HotQCD:2014kol} (blue triangles). The DQPM-CP results are in agreement with the lattice QCD predictions within the estimated errors.
Figure \ref{fig:2d_cs_muB} (b) shows the speed of sound squared $c_s^2$ from the DQPM-CP as a function of the temperature for a wide range of baryon chemical potentials, including the region of the CEP. At high temperatures, values of $c_s^2$ are approaching the limit of a non-interacting gas of massless quarks and gluons (SB limit, black dash-dotted line) $c_s^2(SB)=1/3$.
When increasing the baryon chemical potential the speed of sound near the transition temperature decreases, 
while at the CEP the speed of sound undergoes a sharp decrease.

The DQPM-CP results for the scaled specific heat $C_V/T^3$ as a function of $T$ are presented in Fig. \ref{fig:2d_cv_muB}. As compared to the speed of sound, the specific heat shows an opposite tendency near the phase transition. For moderate values of the baryon chemical potential $\mu_B$ the scaled specific heat increases moderately with decreasing temperature. As it approaches the CEP, $C_V/T^3$ diverges as a function of T, which is  consistent with the expectations for a second-order phase transition. 

The $T$-dependence of the specific heat for $\mu_B=0.96$ GeV near the CEP enables us to estimate the value of the critical exponent for $T>T_{CEP}$:
\begin{equation}
  \mathrm{ln}(C_V)= -\alpha \cdot \mathrm{ln} (T - T_{CEP})+ const. 
\end{equation}
For the presented parametrization of the coupling constant we obtain the following values: $\alpha=0.63\pm 0.02$ and $const=- 5.48\pm 0.01$. The value of the critical exponent $\alpha$ is in agreement with the predictions from the PNJL model for $T>T_{CEP}$ $\alpha_{PNJL}=0.68\pm0.01$ \cite{Costa:2009ae} and the expectations from the universality argument $\alpha=2/3$ in Ref. \cite{Hatta:2002sj}.

In the case of QCD with finite quark masses both the chiral and center symmetries are explicitly broken. What remains is the $Z(2)$ sign symmetry of the order parameter of the chiral phase transition. Therefore it has been assumed that the CEP of QCD with finite quark masses belongs to the three-dimensional $Z(2)$ Ising universality class \cite{Pisarski:1983ms,Berges:1998rc,deForcrand:2003vyj}. The corresponding critical exponent in the $Z(2)$ universality class is $\alpha\approx0.11$ \cite{Pelissetto:2002549}. However, it is known that the critical exponents in the PNJL(NJL) model and $Z(2)$ universality class differ \cite{Sasaki:2007qh,Costa:2007ie,Costa:2009ae,Lu:2015naa,Du:2015psa}.

To explore the high density region, it is essential for effective models to consider isentropic trajectories for which the ratio of entropy to baryon number is held constant. The isentropic trajectories correspond to the ideal hydrodynamical expansion of QGP matter, created in the HICs. When dissipative effects, which can be described by the viscosities and diffusion coefficients, become important, the trajectories are modified \cite{Dore:2020jye,Du:2021zqz}. The presence of the CEP can affect the isentropic trajectories, since the entropy density and baryon density undergo a rapid change as the phase transition is approached. It is supposed that the CEP acts as an attractor of isentropic trajectories \cite{Nonaka:2004pg}. Moreover, a different choice for the strange quark chemical potential affects the trajectories as well. Therefore we compare the resulting isentropic trajectories for two setups of the strange chemical potential.

Figure \ref{fig:isentropes} displays the isentropic trajectories from the DQPM-CP for (a) $\mu_q=\mu_s=\mu_u=\mu_B/3$ and (b) $\mu_s=0,\mu_u=\mu_d=\mu_B/3$ in the phase diagram. Comparing (a) to (b) one can clearly see that the trajectories for the zero strange quark chemical potential are shifted towards higher $\mu_B$ values. In the case of vanishing chemical potential of the strange quark $\mu_s=0,\mu_u=\mu_d=\mu_B/3$, the entropy density, which has also contributions from the light (anti-)quarks and gluons, is less affected then the baryon density. Therefore, for finite $\mu_B>0$ and $\mu_s=0$ the ratio $s/n_B$ is larger than in the case of a symmetric setup $\mu_s=\mu_u=\mu_B/3$ and the value of the baryon density decreases faster than the entropy density.
This observation is in agreement with previous studies of the PNJL model \cite{Motta:2020cbr} and the results from Ref. \cite{Karthein:2021nxe,Parotto:2018pwx}, where the lQCD EoS from the WB collaboration \cite{Borsanyi:2012cr,Borsanyi:2013bia,Gunther:2017sxn} with a critical point in the 3D Ising model universality class is considered for moderate baryon chemical potentials $\mu_B\leq0.45$ GeV. Thus, a 'critical' trajectory, which goes through the CEP, for (a) corresponds to  $s/n_B\approx 13.35$, for (b) corresponds to $s/n_B\approx 15$. The comparison of isentropic trajectories in a vicinity of the CEP is presented in Fig. \ref{fig:isentropes} (c).
In the vicinity of the CEP, the trajectories with $s/n_B=15,13.35,12$ shown in Fig. \ref{fig:isentropes}(c) are focussed to the critical endpoint.

\section{\label{sec4}Transport coefficients}

We continue to investigate the transport properties of QGP matter using the DQPM-CP. We consider the specific shear $\eta/s$ and bulk $\zeta/s$ viscosities, the ratio of electric $\sigma_{QQ}/T$, baryon  $\sigma_{BB}/T$ and strange $\sigma_{SS}/T$ conductivities to temperature. At vanishing baryon chemical potential the DQPM-CP model equals the DQPM, therefore one can find the comparison of DQPM transport coefficients at $\mu_B=0$ with the recent results from various approaches in previous papers \cite{Moreau:2019vhw,Soloveva:2020hpr,Soloveva:2019xph}. 

All transport coefficients  are calculated within the Relaxation Time Approximation (RTA) of the Boltzmann equation.
In the relaxation time approximation (in first order in the deviation from equilibrium) the collision term is given by \cite{Anderson1974RTA}
	 	\begin{align}
		\sum\limits_{j \, = \, 1}^{N_{\text{species}}} \mathcal{C}_{ij}^{(1)}[f_{i}] = -\frac{E_{i}}{\tau_{i}} \left( f_{i} - f^{(0)}_{i}\right) = -\frac{E_{i}}{\tau_{i}} f^{(1)}_{i} + \mathcal{O}(\mathrm{Kn}^2) ,
		\label{eq:RTA}
	\end{align}
where $\tau_i$ is the relaxation time in the heat bath rest system for the particle species $'i'$, $\mathrm{Kn}\sim l_{micro}/L_{macro}$  is the Knudsen number which denotes  the  ratio  between the relevant microscopic scale (mean free path) over the characteristic length scale of the system. 
The equilibrium state of the system is described by the Bose-Einstein and Fermi-Dirac distribution functions
	\begin{align}
	f_{i}^{(0)}(E_i,T,\mu_i) = \frac{ 1 }{\exp \left( (E_i - \mu_i)/T\right) - a_i }  , \label{eq:Equilibrium}
	\end{align}
where $\mu_{i}$ is the quark chemical potential, $E_i=\sqrt{\mathbf{p}_i^2+m_i^2}$ is the on-shell quark/gluon energy, $a_i\equiv +1 \text{(gluons)}, -1  \text{((anti-)quarks)}$.
In Eq. (\ref{eq:RTA}) $f^{(1)}_{i}(x,k,t)$ contains $\delta f_{i}(x,k,t)$, which is  the nonequilibrium part to first order in gradients. \\
The first step in the calculation of the transport coefficients within the RTA framework is the estimation of relaxation times, which are supposed to depend on the momentum of the partons, on the temperature and on the baryon chemical potential. \\
The momentum dependent relaxation time can be expressed through the on-shell 
interaction rate in the rest system of the medium, in which the incoming quark has a four-momentum $P_i=(E_i,\mathbf{p}_i)$ :
 \begin{align}
&\tau^{-1}_i(p_i,T,\mu_q  )= \Gamma_{i}(p_i,T,\mu_q )  \label{eq:gamma_on} \\
& = \frac{1}{2E_i} \sum_{j=q,\bar{q},g} \frac{1}{1+\delta_{cd}} \int \frac{\mathrm{d}^3p_j}{(2\pi)^3 2E_j} d_q f^{(0)}_j(E_j,T,\mu_q)  \nonumber \\
&  \times \int \frac{\mathrm{d}^3p_c}{(2\pi)^3 2E_c}  \int \frac{\mathrm{d}^3p_d}{(2\pi)^3 2E_d} |\bar{\mathcal{M}}|^2 (p_i,p_j,p_c,p_d)\ \nonumber \\  
&  \times (2\pi)^4 \delta^{(4)}\left(p_i + p_j -p_c -p_d \right) (1-f^{(0)}_{c}) (1-f^{(0)}_{d}), \nonumber
 \end{align}

 where $|\bar{\mathcal{M}}|^2$ denotes the matrix element squared averaged over the color and spin of the incoming partons, and summed over those of the final partons. The $|\bar{\mathcal{M}}|^2$ is calculated by the use of the effective coupling and propagators in leading order (for further details see \cite{Moreau:2019vhw}).
 The notation $\sum_{j=q,\bar{q},g}$ includes the contribution
from all possible partons which in our case are the gluons
and the (anti-)quarks of three different flavors ($u$, $d$, $s$).
The quark relaxation time is expected to become very large near CEP, since the correlation length increases rapidly close to the CEP.

\subsection{Specific viscosities}

\begin{figure}[!ht]
\begin{minipage}[h]{1\linewidth}
\center{\includegraphics[width=1.0\linewidth]{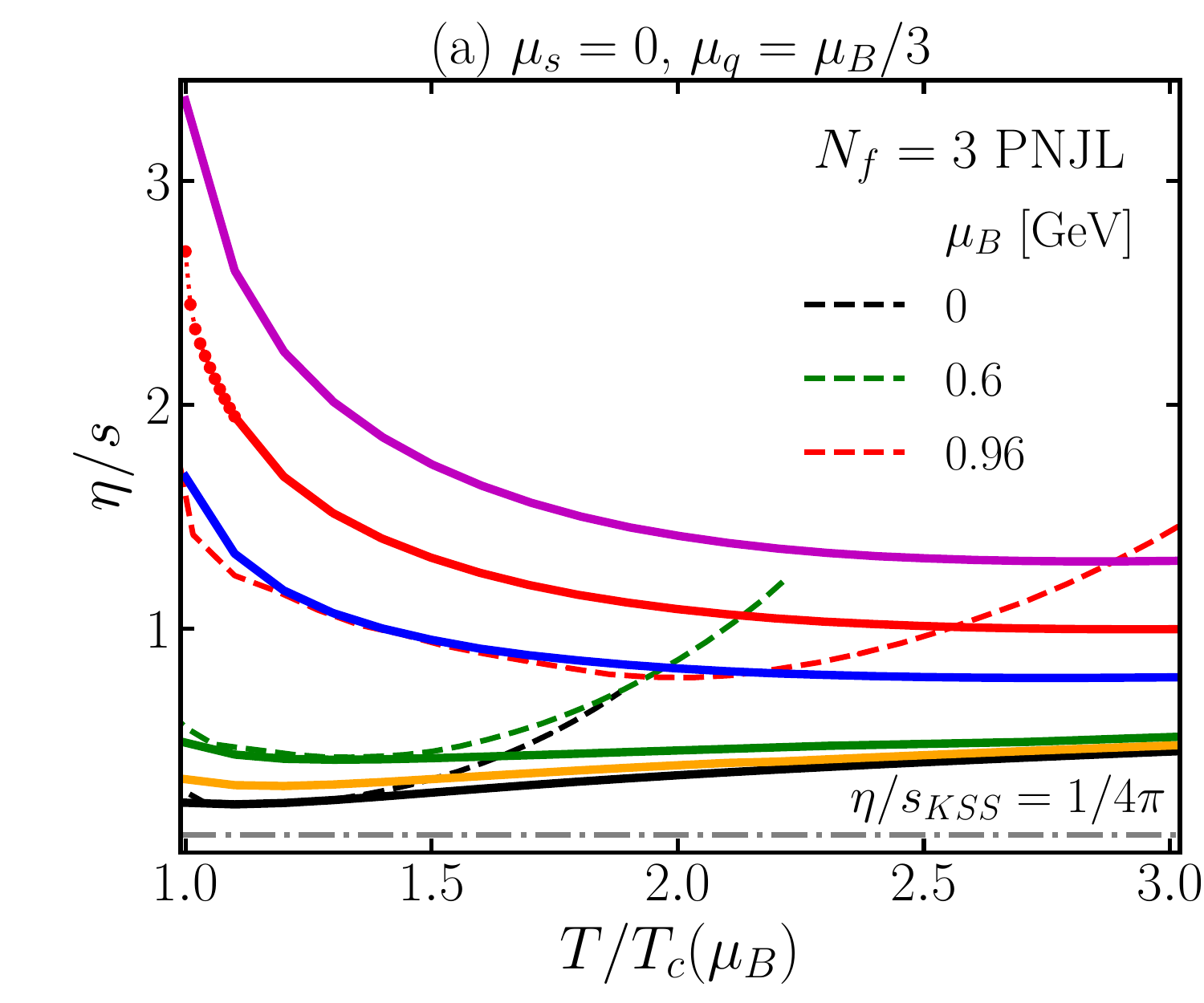}}
\end{minipage}
\begin{minipage}[h]{1\linewidth}
\center{\includegraphics[width=1.0\linewidth]{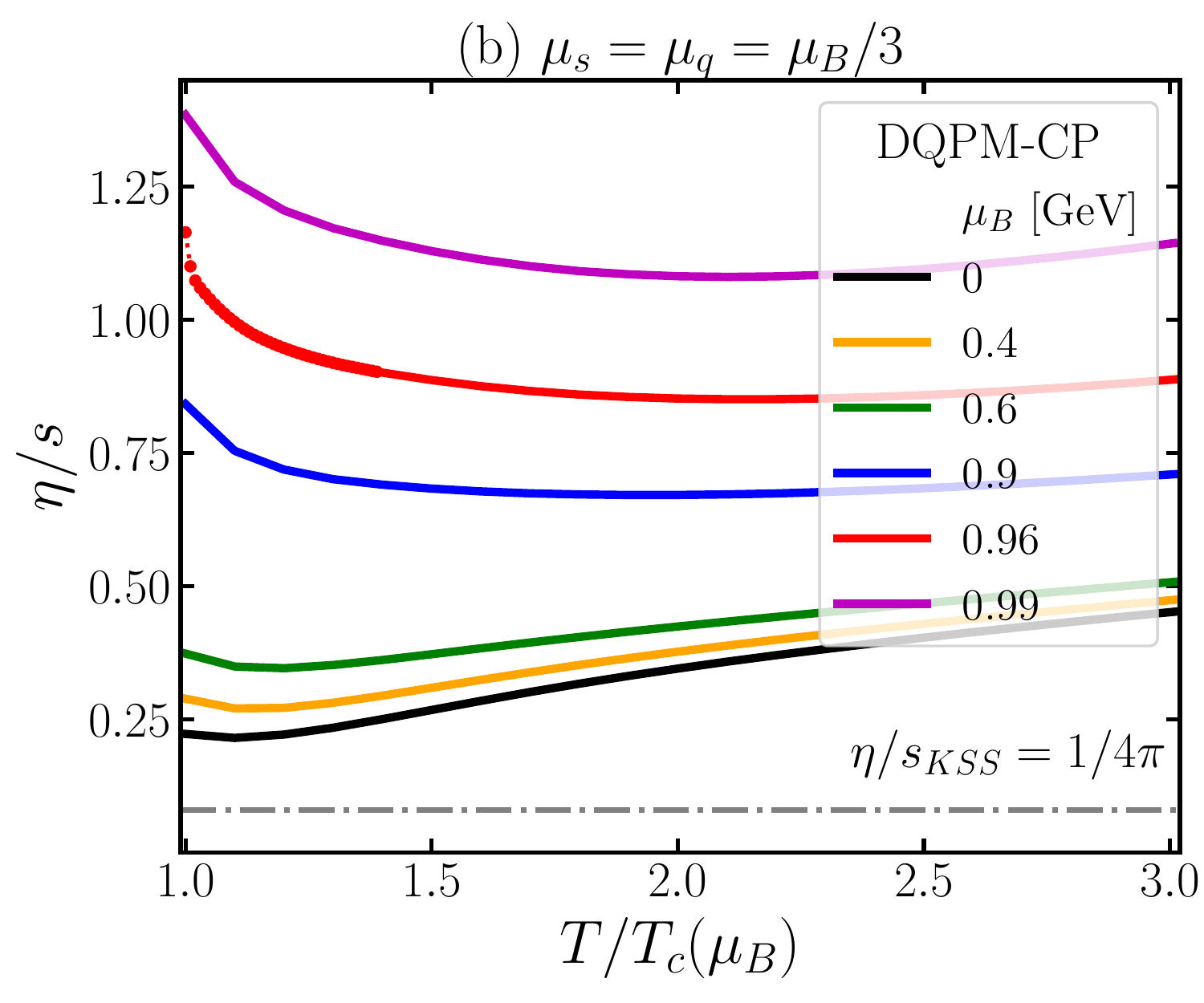}}
\end{minipage}
\caption{\label{fig:etas_mu}Specific shear $\eta/s$ viscosity from the DQPM-CP (solid lines) for two setups of strange chemical potential: (a) ($\mu_s=0,\mu_u=\mu_B/3$) and (b) ($\mu_s=\mu_u=\mu_B/3$) as a function of the scaled temperature $T/T_c$ for various $\mu_B\geq 0$. We compare to the RTA estimates from the $N_f=3$ PNJL model (dashed lines) \cite{Soloveva:2020hpr}. The grey dashed-dotted line demonstrates the Kovtun-Son-Starinets bound \cite{Kovtun:2004de} $(\eta/s)_{\rm{KSS}} = 1/(4\pi)$.}
\end{figure}
We start with the most common  transport coefficients for the hydrodynamical simulations - the shear and bulk viscosity. The viscosities of the QCD matter have been studied within a variety models in the confined and the deconfined phases. The shear viscosity reveals the strength of the interaction inside the QCD medium, in particular within the kinetic theory it can be related to the hadron or parton interaction rates, which is a challenge to evaluate on the basis of first principles. A plethora of theoretical model predictions show that the temperature dependence of the QCD shear viscosity over entropy density $\eta/s$ is qualitatively different for the two phases. Starting from the hadronic phase below the phase transition $T<T_c$, $\eta/s$ monotonically decreases with $T$ since the system is dominated by pions with weaker interactions at lower $T$. While above the phase transition  $T>T_c$, $\eta/s$  increases with temperature because the interaction attenuates at high T. Approaching the phase transition from hadronic to the QGP phase at vanishing chemical potential, $\eta/s$ has a wide dip followed by an increase with temperature. A similar property of the  temperature dependence of the specific shear viscosity $\eta/s$ is seen for other fluids such as $H_2O$, $He$ and $N_2$ \cite{codata:1989,Tournier_etasN2:2008,Brazhkin:2012PRE}. \\
The specific bulk viscosity of the QGP matter is predicted to be low, yet it is expected to be finite near the phase transition \cite{Ryu:2015vwa}. The presence of the bulk viscosity reduces the speed of the fluid radial expansion and hence affects the mean momentum of the produced particles.
For conformal fluids, the bulk viscosity is known to be identically zero, and the deconfined QCD medium is expected to adopt a conformal behaviour in the high-energy or temperature regime. Nevertheless, the lQCD results on the enhanced trace anomaly close to $T_c$ have shown that it is probably not the case for the deconfined QCD medium in the vicinity of the phase transition.

\begin{figure}[!h]
\begin{minipage}[h]{1\linewidth}
\center{\includegraphics[width=1.0\linewidth]{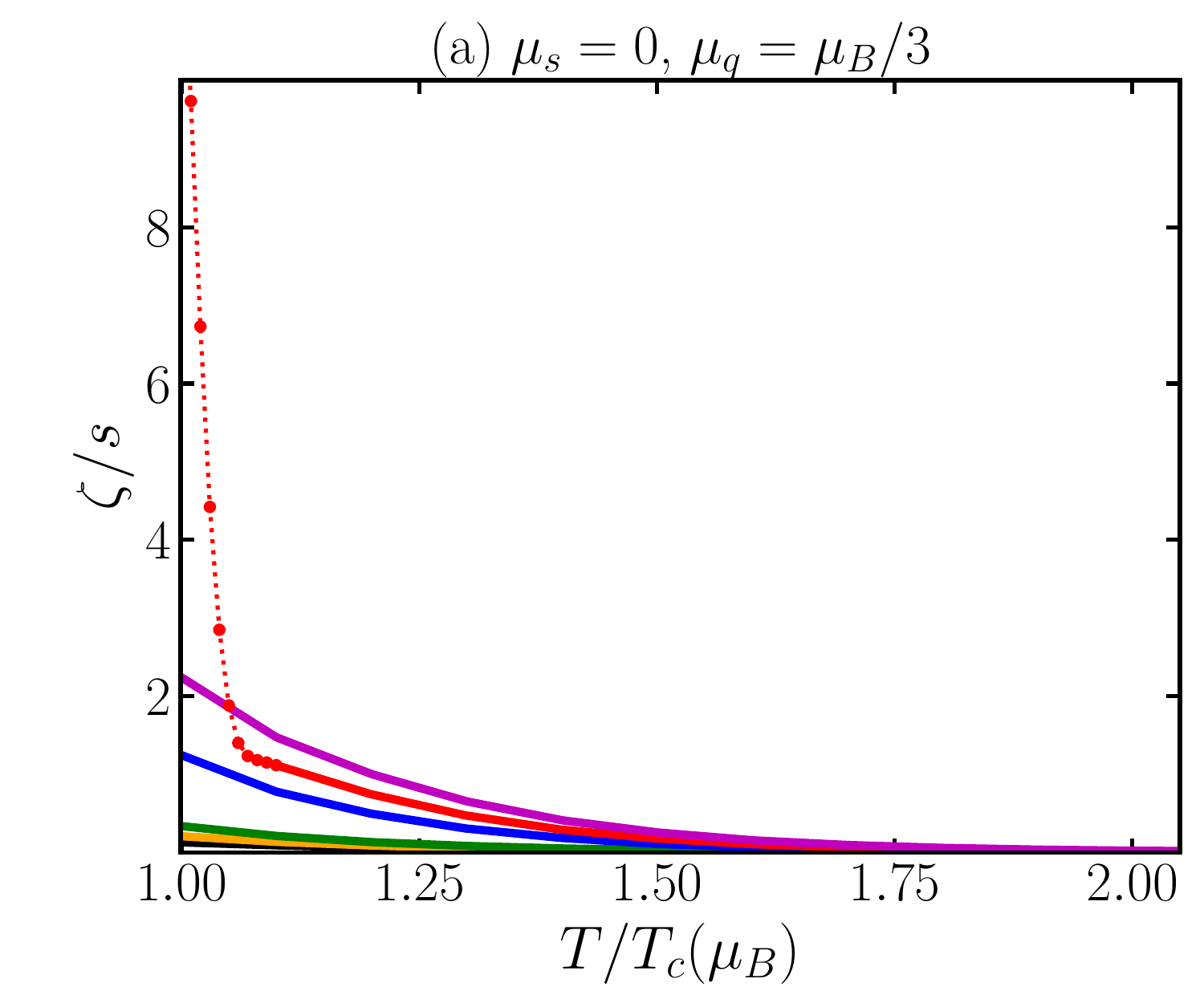}}
\end{minipage}
\begin{minipage}[h]{1\linewidth}
\center{\includegraphics[width=1.0\linewidth]{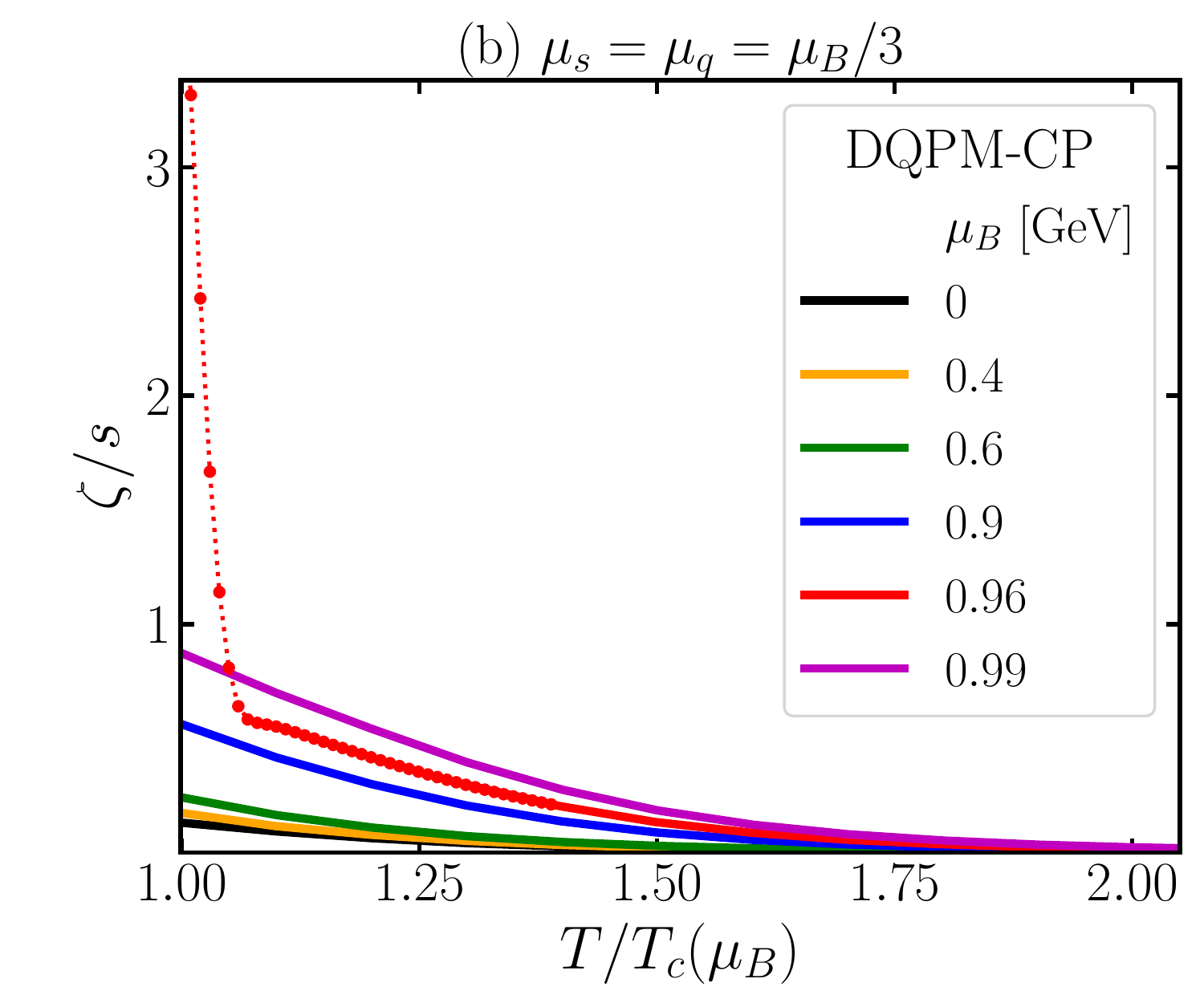}}
\end{minipage}
\caption{\label{fig:zetas_mu}Specific bulk $\zeta/s$ viscosity from the DQPM-CP (solid lines) for two setups of strange chemical potential: (a) ($\mu_s=0,\mu_u=\mu_B/3$) and (b) ($\mu_s=\mu_u=\mu_B/3$) as a function of the scaled temperature $T/T_c$ for various $\mu_B\geq 0$.}
\end{figure}
 
The shear and bulk viscosity for quasiparticles with medium-dependent masses $m_i(T,\mu_q)$ can be derived using the Boltzmann  equation in the RTA \cite{Chakraborty:2010fr} through the relaxation time:
\begin{align}
\eta(T,\mu_q)  = \frac{1}{15T} \sum_{i=q,\bar{q},g} \int \frac{d^3p}{(2\pi)^3} \frac{\mathbf{p}^4}{E_i^2}   \tau_i(\mathbf{p},T,\mu_q) \nonumber \\ 
\times   d_i  (1 \pm f_i) f_i , 
\label{eq:eta_RTA}
\end{align}

\begin{align}
\zeta(T,\mu_q)= \frac{1}{9T} \sum_{i=q,\bar{q},g}\int \frac{d^3p}{(2\pi)^3}    
  \label{eq:zeta_on} \tau_i(\mathbf{p},T,\mu_q)  \nonumber \\
\frac{ d_i (1 \pm f_i) f_i  }{E_i^2}
\left[\mathbf{p}^2-3c_s^2(E_i^2-T^2\frac{dm_i^2}{dT^2})\right]^2 ,
\end{align}
where $q(\bar{q})=u,d,s(\bar{u},\bar{d},\bar{s})$, $d_q = 2N_c = 6$ and $d_g = 2(N_c^2-1) = 16$ are the degeneracy factors for spin and color for quarks and  gluons, respectively, $\tau_i$ are the relaxation times. $c_s$ is the speed of sound for a fixed $\mu_B$ given by Eq. (\ref{eq:cs2}), $\frac{dm_i^2}{dT^2}$ is the derivative of the effective masses.
As it was shown in the previous studies \cite{Sasaki:2008um,Berrehrah:2015ywa,RMarty:2013xph,Moreau:2019vhw,Soloveva:2019xph}
in case of the medium dependent masses, the viscosities display a pronounced temperature behaviour. At vanishing baryon chemical potential we found previously that the DQPM results for specific shear and bulk viscosity  \cite{Moreau:2019vhw, Soloveva:2020hpr} are very close to the predictions from the gluodynamic lQCD calculations \cite{Astrakhantsev:2017nrs,Astrakhantsev:2018oue}.

 For moderate values of the baryon chemical potential, the specific shear viscosity $\eta/s$  of the QGP matter increases with temperature, while the specific bulk viscosity $\zeta/s$ decreases with temperature, independent of baryon chemical potential. However, the T dependence of $\eta/s$ near the phase transition changes with increasing $\mu_B$. Fig. \ref{fig:etas_mu} shows the DQPM-CP results for $\eta/s$ as a function of scaled temperature $T/T_c(\mu_B)$, for different $\mu_B$ values. The specific shear viscosity for $\mu_B=0$ (red line) shows a dip followed by an increase with temperature while for a $\mu_B\geq0.9$ GeV $\eta/s$ decreases with increasing temperature near the phase transition $(T\leq 2T_c)$.
The parton relaxation time $\tau_i$ decreases with increasing temperature at lower T, and remains approximately constant at high T for moderate values of chemical potential $\mu_B\geq 0.6$ GeV. Therefore the shear viscosity $\eta \sim T^4$, while the entropy density grows as $s \sim T^3$. Thus, in the high temperature region the ratio $\eta/s$ increases as $\sim T$.
It is important to note that transport coefficients rely on the microscopic properties of the degrees of freedom. 
While a variety of the models can reproduce the lQCD results of basic thermodynamic observables, the transport coefficients differ between the models. 

Here we compare results of the specific shear viscosity and later of the electric conductivity for non-zero baryon or quark chemical potential with the RTA results from the PNJL model for $\mu_s=0,\mu_u=\mu_B/3$ in Fig. \ref{fig:etas_mu} (a). The specific shear viscosity results from the DQPM-CP (solid lines) for $\mu_B=0,0.6$ GeV agree well with the predictions of the PNJL model (dashed lines) in the vicinity of the phase transition for temperatures $T\leq 1.5T_c$. The DQPM-CP results for $\eta/s$ at $\mu_B=0.96$ GeV for $T\leq 2T_c$ is higher than from the PNJL model, while the temperature dependence is similar.
The discrepancy between the results is caused by the different treatment of the gluonic degrees of freedom , which has a pronounced critical behaviour of the thermal masses in the DQPM-CP model.
Increasing the baryon chemical potential, one can see not only an increase in magnitude but also a change in the $T$-dependence of $\eta/s$ and $\zeta/s$ as shown in Fig. \ref{fig:etas_mu}.
\begin{figure}[!ht]
\centering
\begin{minipage}[h]{1\linewidth}
\center{\includegraphics[width=0.98\linewidth]{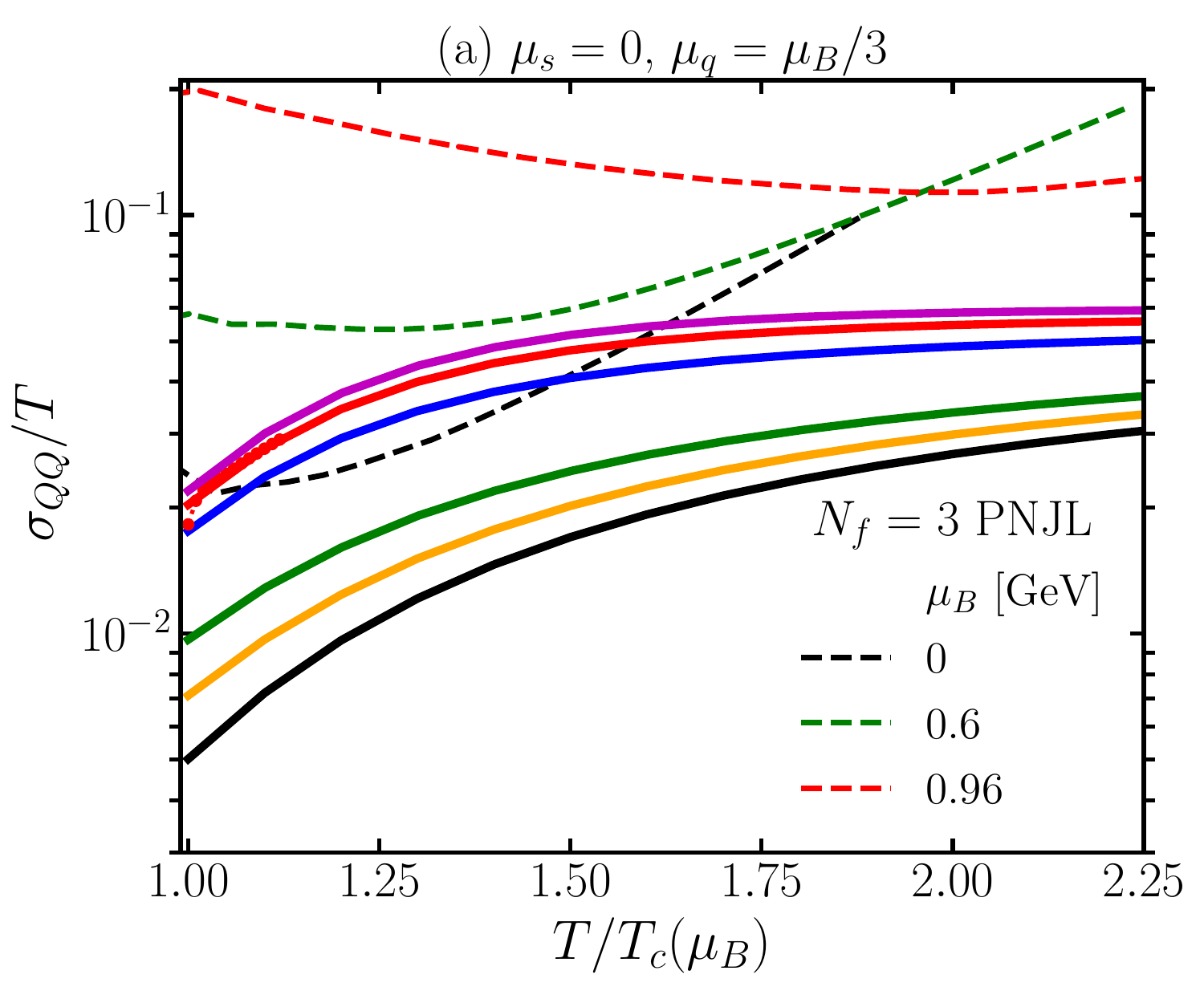}}
\end{minipage}
\begin{minipage}[h]{1\linewidth}
\center{\includegraphics[width=0.98\linewidth]{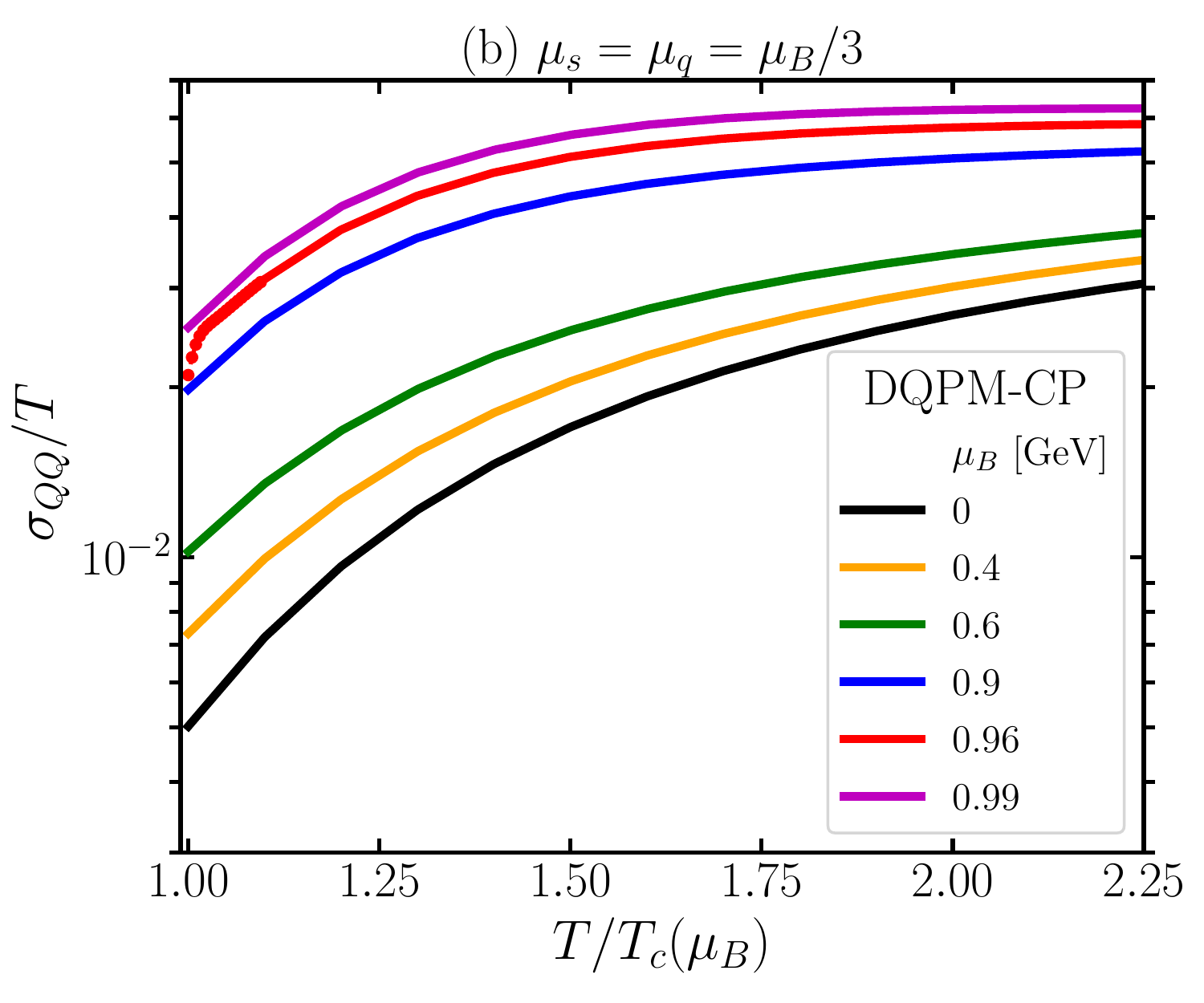}}
\end{minipage}
\caption{\label{fig:musigmaqq}  Scaled electric conductivity as a function of the scaled temperature $T/T_c$ from the DQPM-CP (solid lines) for two setups of strange chemical potential: (a) ($\mu_s=0,\mu_u=\mu_B/3$) and (b) ($\mu_s=\mu_u=\mu_B/3$)  as a function of the scaled temperature $T/T_c$ for various $\mu_B\geq 0$. For ($\mu_s=0,\mu_u=\mu_B/3$) we compare $\sigma_{QQ}/T$ from DQPM-CP to the RTA estimates from the $N_f=3$ PNJL model (dashed lines) \cite{Soloveva:2020hpr}.}
\end{figure}
In particular, in the vicinity of the phase transition $T<1.5\ T_c$ for moderate values of $\mu_q$ the specific shear viscosity shows a dip after the phase transition, which is vanishing at high values of $\mu_B$ as can be seen in Fig.~\ref{fig:etas_mu}. \\
As pointed out in Refs. \cite{Son:2004iv,Karsch:2007jc,Moore:2008ws,Sasaki:2008um} in the vicinity of the CEP, the divergences of bulk and shear viscosities of the QCD matter are determined by the dynamic and the static critical exponents.
 The dynamical universality class of the QCD critical endpoint is argued  to be that of the H-model \cite{Son:2004iv,Fujii:2004jt} according to the classification of dynamical critical phenomena by Hohenberg and Halperin  \cite{Hohenberg:1977ym}. Whereas in the vicinity of the CEP the shear viscosity has a mild divergence in the critical region, the bulk viscosity has a more pronounced divergence \cite{Kadanoff:1968zz,Hohenberg:1977ym,Son:2004iv}: $\eta \sim \xi_T ^{Z_\eta} (Z_\eta \approx 1/19)$,  $\zeta \sim \xi_T^{Z_\zeta}(Z_\zeta \approx 3)$. The thermal correlation length is controlled by the static critical exponent $\xi_T \sim t^{-\nu}$, $t=\frac{T-T_c}{T_c}$, with $\nu$ being the static critical exponent. Using the hyperscaling relation \cite{Goldenfeld:1992qy} for the static critical exponents we can estimate $\nu$:
 \begin{equation}
  2-\alpha=d \nu,   
 \end{equation}
 where $d=3$ denotes the number of the spatial dimensions, $\alpha \approx 0.63$. We obtain $\nu \approx 0.46$. Taking into account the dynamical and static exponents the divergence of the bulk viscosity is assumed to be $\zeta\approx t^{-Z_\zeta\nu+\alpha}$ \cite{Kadanoff:1968zz,Onuki:1997lg,Moore:2008ws}.
 
Here we consider small deviations from equilibrium where the quark relaxation times are not large: $\tau_q$ is about  $4.5 - 2.5$  fm/c for the temperature range $T_c<T\leq 2T_c$, so that the slight divergence of the transport coefficients near the CEP is determined by the static exponents.
The specific shear and bulk viscosities from the DQPM-CP increase rapidly when approaching the critical endpoint from the partonic phase. However, the increase near the CEP is more pronounced for the specific bulk viscosity which rises by a factor of five, while the specific shear viscosity rises only by $\leq$ 10\% for the same temperature range $1.07-1.01$ $T/T_c$. The increase of $\zeta/s$ is related to the rapid decrease in the speed of sound and corresponds to the static critical exponents, that affects the bulk viscosity. In terms of heavy-ion collisions observables, this increase in the bulk viscosity is expected to show up as the decrease of average transverse momentum of produced particles as well as in an increase of the charged particle multiplicity per unit momentum rapidity  \cite{Karsch:2007jc,Ryu:2015vwa}. However, this has to  be checked by a transport  simulations or by a hydrodynamical simulation of the expanding QGP. Such a substantial increase of the charged particle and net-baryon multiplicities per unit momentum rapidity due to the enhancement of the bulk viscosity near the CEP has been observed in a longitudinally expanding 1 + 1
dimensional causal relativistic hydrodynamical evolution at non-zero baryon density \cite{Monnai:2016kud}. 

We note that the specific bulk and shear viscosities have been considered near the CEP and the 1st order phase transition for the $N_f = 2$ NJL model in the previous study \cite{Sasaki:2008um}. We found good qualitative agreement for the $T$-dependence of the shear and bulk viscosity of the NJL model from Ref. \cite{Sasaki:2008um}, while the numerical values differ due to the different quark relaxation times and the absence of gluonic degrees of freedom in case of the NJL model.

\subsection{Electric, baryon and strange conductivities}
\begin{figure}[!ht]
\centering
\begin{minipage}[h]{1\linewidth}
\center{\includegraphics[width=0.98\linewidth]{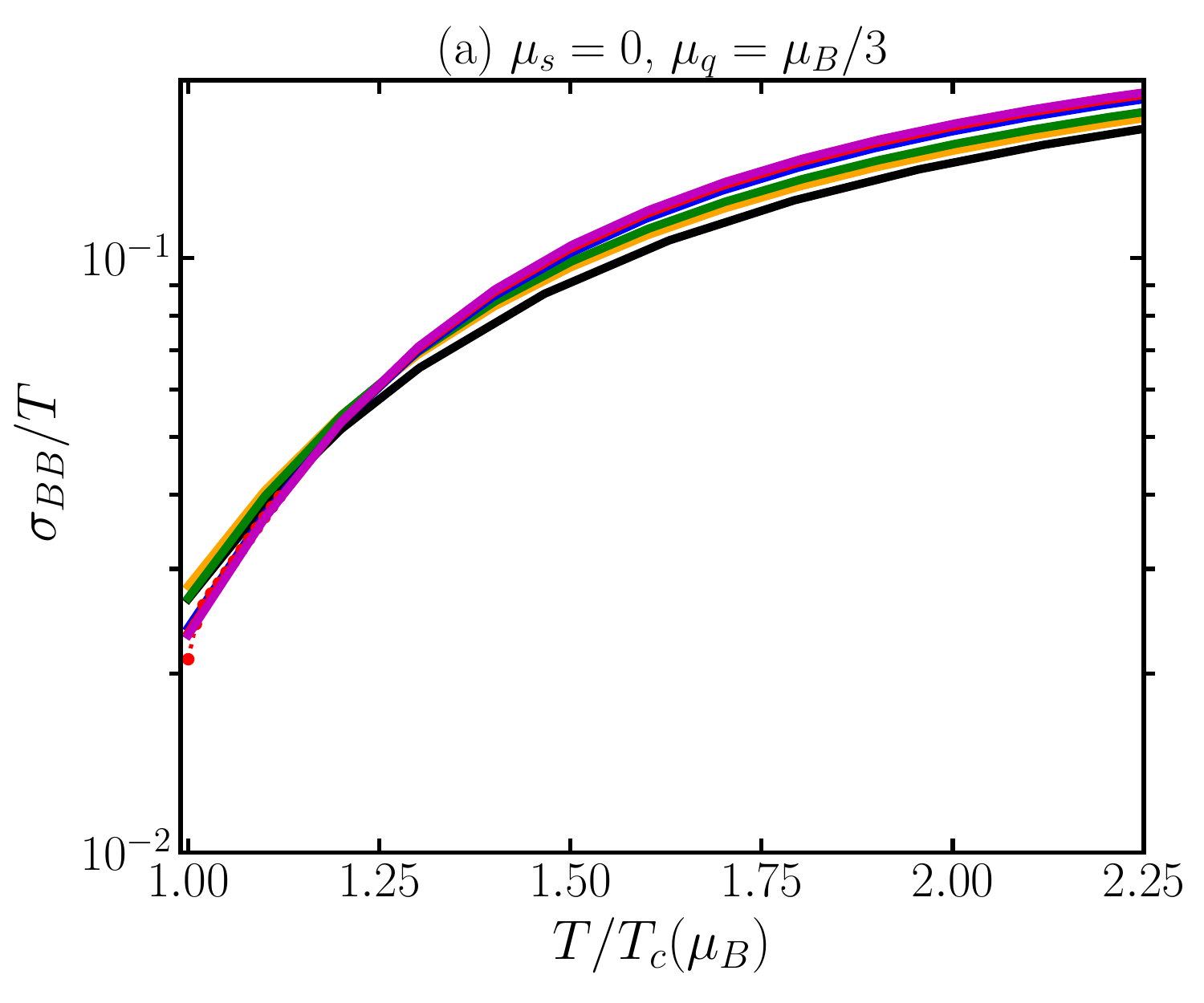}}
\end{minipage}
\begin{minipage}[h]{1\linewidth}
\center{\includegraphics[width=0.98\linewidth]{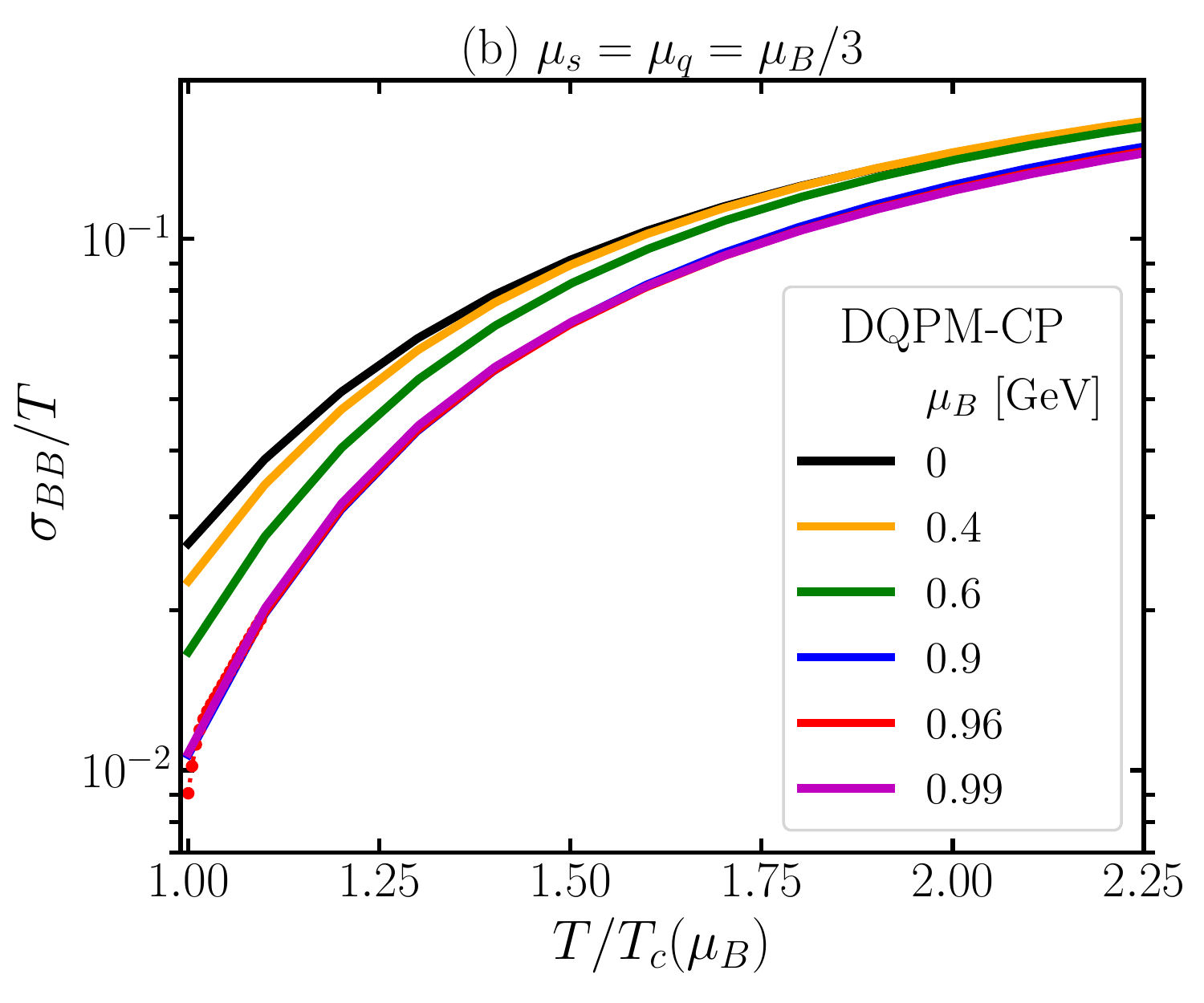}}
\end{minipage}
\caption{\label{fig:musigmabb} Scaled baryon conductivity as a function of scaled temperature $T/T_c$ from the DQPM-CP (solid lines) for two setups of strange chemical potential: (a) ($\mu_s=0,\mu_u=\mu_B/3$) and (b) ($\mu_s=\mu_u=\mu_B/3$) as a function of the scaled temperature $T/T_c$ for various $\mu_B\geq 0$.}
\end{figure}
\begin{figure}[!ht]
\centering
\begin{minipage}[h]{1\linewidth}
\center{\includegraphics[width=0.98\linewidth]{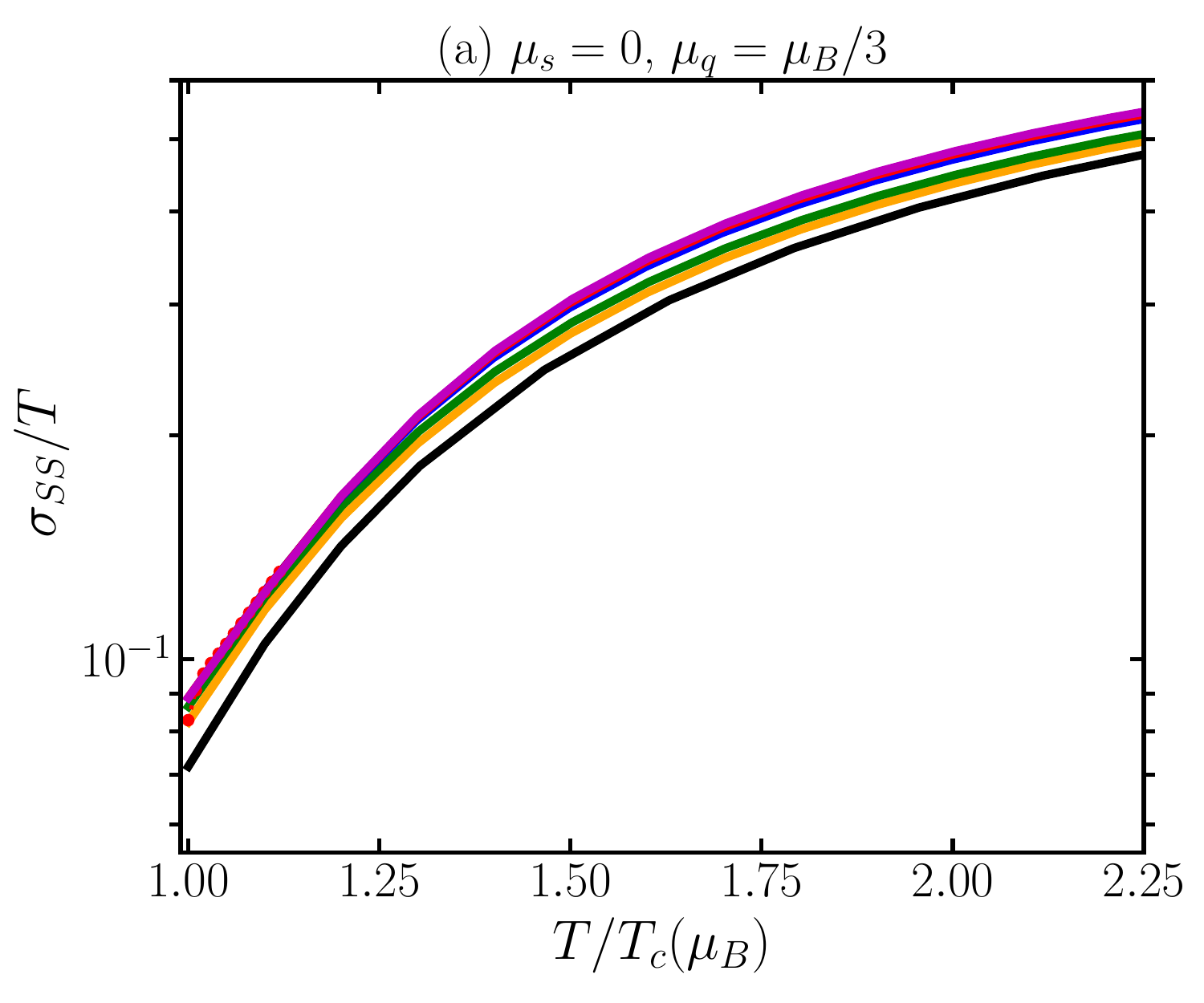}}
\end{minipage}
\begin{minipage}[h]{1\linewidth}
\center{\includegraphics[width=0.98\linewidth]{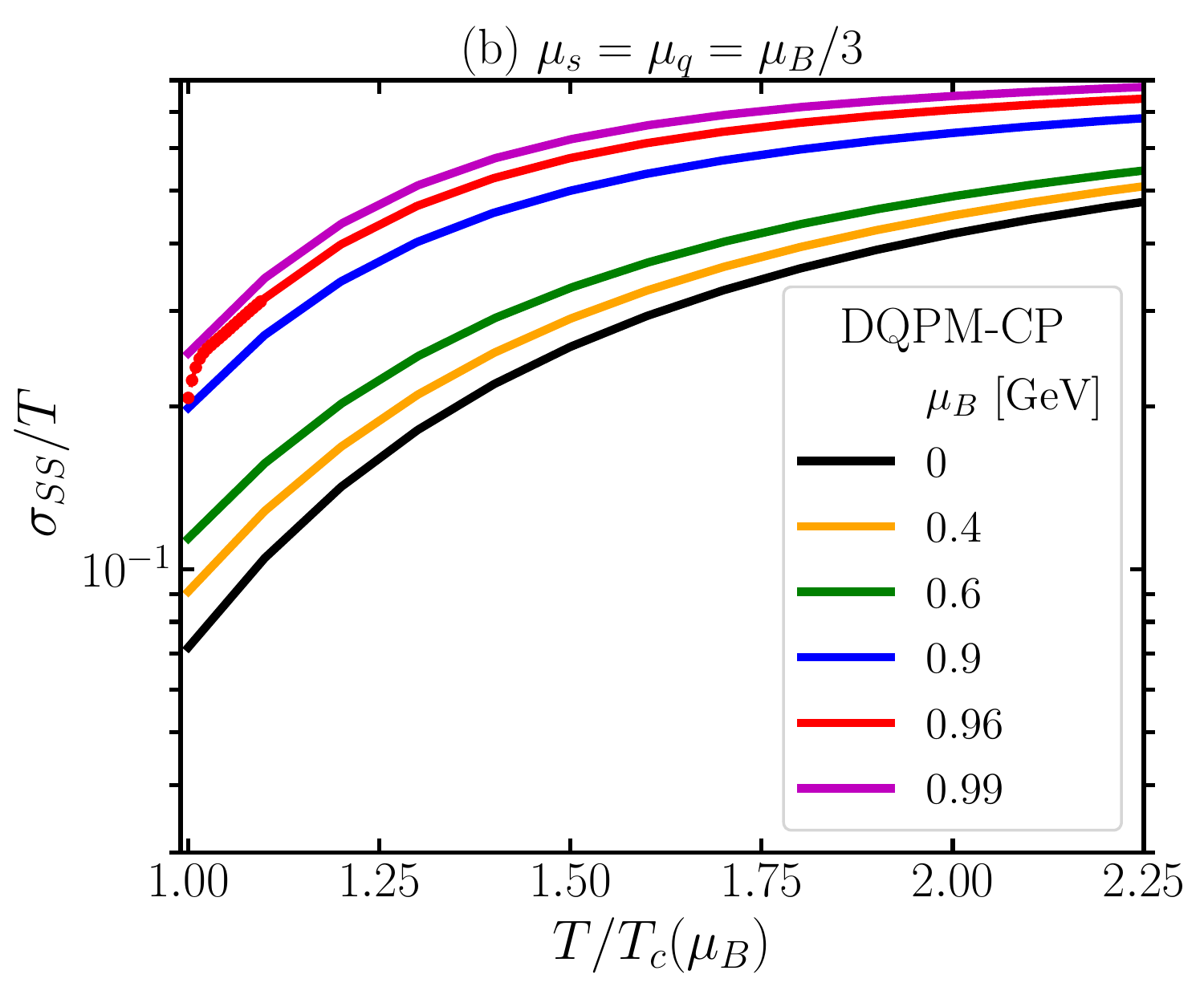}}
\end{minipage}
\caption{\label{fig:musigmass} Scaled strange conductivity as a function of scaled temperature $T/T_c$ from the DQPM-CP (solid lines) for two setups of strange chemical potential: (a) ($\mu_s=0,\mu_u=\mu_B/3$) and (b) ($\mu_s=\mu_u=\mu_B/3$) as a function of the scaled temperature $T/T_c$ for various $\mu_B\geq 0$.}
\end{figure}

In the region of the high net baryon density it is important to take into account the diffusion of conserved charges, i.e. electric, baryon and strange charges, from higher density regions to lower density regions. The transport coefficient, which characterizes the diffusion, is the diffusion coefficient $\kappa_{q}$ or the conductivity $\sigma_{q}=\kappa_{q}/T$ of the conserved charge $q$. Furthermore, since the quarks carry multiple conserved charges, one needs to consider additionally non-diagonal conductivities for two conserved charges $qq'$ -- $\sigma_{qq'}$. Conductivities $\sigma_{qq'}$ for a quasiparticles can be expressed in the RTA \cite{Fotakis:2019nbq} as:
	\begin{align}
		\sigma{qq^\prime}(T,\mu_q) = \frac{1}{3T}  \sum\limits_{i=q,\bar{q}} \, q_i \int \frac{d^3p}{(2\pi)^3} \frac{\mathbf{p}^2}{E_i^2}  \, \tau_i(\mathbf{p},T,\mu_q) \nonumber \\
		\times \left( \frac{E_{i} n_{q^\prime}}{\epsilon + p} - q^\prime_i \right) d_i (1 \pm f_i) f_i  . \label{eq:DiffMatrRTA}
	\end{align}
Let us consider first the diagonal conductivities for electric, baryon and strange charge.
The DQPM-CP results for $\sigma_{QQ}/T$,  $\sigma_{BB}/T$ and  $\sigma_{SS}/T$ are shown in Figs. \ref{fig:musigmaqq}, \ref{fig:musigmabb}, \ref{fig:musigmass} as a function of the scaled temperature $T/T_c$ for two setups of the strange quark chemical potential.
The scaled electric, strange and baryon conductivity have a similar temperature dependence: at high $T$ the ratios increase with temperature increase as $\sim T$ which is mainly due to the quark density increasing with temperature.
The most prominent difference between the conductivities is the $\mu_B$-dependence, which is shown in Figs. \ref{fig:musigmaqq}, \ref{fig:musigmabb}, \ref{fig:musigmass}: the electric and strange conductivities increase with $\mu_B$, while the baryon conductivity decreases with $\mu_B$ for the symmetric setup $\mu_s=\mu_u=\mu_B/3$.   With the increase of baryon chemical potential the net baryon density increases, which influences the baryon conductivity. A similar trend for the $\sigma_{QQ}/T$,  $\sigma_{BB}/T$ and  $\sigma_{SS}/T$ at moderate values of baryon chemical potential $\mu_B\leq 0.4$ GeV has been observed in the non-conformal Einstein-Maxwell-Dilaton (EMD) holographic model \cite{Rougemont:2017tlu}.
Futhermore, we compare the $\mu_B$ dependencies of the scaled conductivities for the two setups of strange quark chemical potential. We have found that in case of vanishing strange quark chemical potential (setup (II)) the scaled conductivities show a much less pronounced of $\mu_B$-dependence for the baryon and strange conductivities, which is expected due to the vanishing net strangeness density $n_S=0$. While the electric conductivity has similar $\mu_B$-dependence for the two settings of strange quark chemical potential.
Near the CEP, the electric conductivity decreases, but as for the PNJL results, there is no pronounced divergence behavior. The same behavior has been found for the baryon and strange conductivities.
~\\

\section{\label{sec5}Conclusions and Outlook}
By extending the phenomenological dynamical quasiparticle model to a wide range of baryon chemical potentials we obtain an EoS, which is in agreement with the lattice data at moderate baryon chemical potentials and can at the same time be extended to the whole ($T$, $\mu_B$) plane. This extension allows for calculating the transport coefficients of the partonic phase.
To mimic the $T$-dependence of the basic thermodynamic observables near the CEP we have adopted the critical behaviour of the effective coupling constant by using the entropy density from the PNJL model near the CEP.  For moderate values of the chemical potential $\mu_B \leq 0.4$ GeV the dependence of the thermodynamic quantities on $\mu_B$ are in agreement with the previous results from the DQPM \cite{Berrehrah:2013mua, Berrehrah:2016vzw,Moreau:2019vhw}.
\begin{itemize}
\item[$\blacktriangleright$]
We presented the results for the thermodynamic observables $p/T^4$, $\epsilon/T^4$, $s/T^3$, as well as for the speed of sound and the specific heat for a wide range of chemical potentials. We have shown that the 'critical' behaviour of the effective coupling affects the thermodynamic observables. Moreover, we have found that the resulting value of the critical exponent $\alpha\approx0.63$ is in good agreement with the predictions of the PNJL model and the expectations from the universality argument $\alpha=2/3$.
\item[$\blacktriangleright$]
To quantify the $\mu_B$-dependence of the bulk observables we have  have studied isentropic trajectories of the deconfined QCD medium described by the DQPM-CP for a wide range of baryon chemical potential, including the vicinity of the CEP. 
\item[$\blacktriangleright$] 
We have evaluated transport properties of the deconfined QCD medium for a wide range of baryon chemical potential within the DQPM-CP: the specific shear $\eta/s$ and bulk $\zeta/s$ viscosity and the ratio of electric $\sigma_{QQ}/T$, baryon  $\sigma_{BB}/T$ and strange $\sigma_{SS}/T$ conductivities to temperature on the basis of the Boltzmann equation in the relaxation time approximation. 
We have found that the resulting $\mu_B$-dependence of $\eta/s$ and $\sigma_{QQ}/T$ for the PNJL model and the DQPM-CP are qualitatively the same in the vicinity of the phase transition, while there is a clear difference in the electric conductivity.
\item[$\blacktriangleright$] 
We have found that the DQPM-CP estimates of the specific bulk viscosity show a rapid increase when approaching the CEP from the high-temperature region originating from the rapid decrease of the speed of sound $c^2_s\rightarrow 0$, whereas for the specific shear viscosity and the $B,Q,S$ conductivities there is only a small enhancement $\leq 10\%$, caused mainly by the `critical' contribution of the effective coupling constant.
\end{itemize}
Although the extracted results for the transport coefficients are model-dependent, the qualitative picture of the $T$ and $\mu_B$ dependence is consistent with expectations from more rigorous approaches. 
Our results can be implemented in  hydrodynamic simulations as well as be employed 
for the partonic phase of  transport approaches. 
~\\
\begin{acknowledgments}
The authors thank Wolfgang Cassing, Juan Torres-Rincon, Claudia Ratti and 
Taesoo Song for useful discussions.
O.S. acknowledge support from the \textquotedblleft Helmholtz Graduate School for Heavy Ion research\textquotedblright. 
O.S. and E.B. acknowledge support by the Deutsche Forschungsgemeinschaft (DFG, German Research Foundation) through the grant CRC-TR 211 'Strong-interaction matter under extreme conditions' - Project number 315477589 - TRR 211. 
This work is supported by the European Union’s Horizon 2020 research and innovation program under grant agreement No 824093 (STRONG-2020) 
and by the COST Action THOR, CA15213.  
Computational resources were provided by the Center for Scientific Computing (CSC) of the Goethe University.
\end{acknowledgments}

\bibliographystyle{apsrev4-1}
\bibliography{paper2021.bib} 

\end{document}